\documentstyle[epsfig,aps,prc]{revtex}

\begin{document}
\title{ Bose-Einstein condensation of correlated atoms in a trap }

\author{Ch. C. Moustakidis and S. E. Massen}

\address{Department of Theoretical Physics,
Aristotle University of Thessaloniki,
GR-54006 Thessaloniki, Greece}
\maketitle

\begin{abstract}
The Bose-Einstein condensation of correlated atoms in a trap is studied
by examining the effect of inter-particle correlations
to one-body properties of atomic systems at zero temperature using
a simplified formula for the correlated two body
density distribution.
Analytical expressions for the density distribution
and rms radius of the atomic
systems are derived using four different expressions of Jastrow type
correlation function. In one case, in addition, the
one-body density matrix, momentum distribution
and kinetic energy are calculated analytically,
while the natural orbitals and natural occupation numbers
are also predicted in this case.
Simple approximate
expressions for the mean square radius and kinetic energy
are also given.

\end{abstract}

\section{Introduction}

The first theoretical prediction of the famous phenomenon known as
Bose-Einstein condensation (BEC) was made in 1924 and 1925 by
Bose \cite{Bose24} and Einstein \cite{Einstein 24-25}, respectively.
In a system of particles obeying Bose-Einstein statistics where the
total number of particles is conserved, there should be a temperature
below which a finite fraction of all the particles condense into the
same one-particle state 
\cite{Griffin95,Dalfano99,Leggett01,Parkins98,Fetter98,Courteille}.
Seventy years later, in a remarkable experiment, Anderson et al.
\cite{Anderson95} have cooled magnetically
trapped $^{87}$Rb gas to nanokelvin temperatures, and observed the
BEC. This discovery has generated a huge amount
of theoretical investigations \cite{Baym96,Dalfano96a,Dalfano96b,%
Wu98,Fabrocini99,Fabrocini01,Fabrocini00,Giorgini99,DuBois 01,Cherny01,%
Ziegler97,Minguzzi00,Naraschewski99,Pitaevski99,Kuklov99,Kavoulakis00,%
Proukakis97,Banerjee01}.

The main feature of the trapped alkali-metal and atomic hydrogen systems
(which obey the Bose-Einstein statistics) is that they are dilute, i.e.,
the average distance among the atoms is much larger than the range
of the interaction. The crucial parameter defining the condition of
diluteness is the gas parameter $\chi=na^3$, where $n$ is the density
of the system
and $a$ is the s-wave scattering length \cite{Fabrocini01}.
There are two ways to bring $\chi$ outside the regime of validity
of the mean field description. The first one consists by increasing
the density, while the second one consists
by changing the effective size of the atoms
\cite{Fabrocini00}. In the present work we follow the second way.

The characteristic dimension of a harmonic oscillator (HO) trap
for $^{87}$Rb is $b=\left(\frac{\hbar}{m
\omega}\right)^{1/2}=(1-2) \times 10^4 \AA$ while the scattering
length lies in the range $85 a_o < a < 140 a_o$, where $a_o=0.5292
\AA$ is the Bohr radius. The atomic density in the trap is
$n\simeq10^{12}-10^{14}$ atoms/cm$^3$ giving an inter-atomic
distance $l=\left(\frac{V}{A}\right)^{1/3}\simeq10^4 \AA$
\cite{DuBois 01}. In this case, the effective atomic size is small
compared both to the trap size and to the inter-atomic distance
ensuring the diluteness of the gas. However, the effects of
inter-particle interactions are of fundamental importance in the
study of the  BEC dilute-gas where the physics should be dominated
by two-body collisions described in terms of the s-wave scattering
length $a$. In the case of positive $a$, it is equivalent to
consider a very dilute  (atomic) system of hard spheres, whose
diameter coincides with the scattering length itself
\cite{Fabrocini99}. The natural starting point for studying the
behavior of those systems is the theory of weakly interacting
bosons which, for inhomogeneous systems, takes the form of the
Gross-Pitaevskii theory \cite{Pitaevskii61,Gross61}. This is a
mean-field approach for the order parameter associated with the
condensate. It provides closed and relatively simple equations for
describing the relevant phenomena associated with BEC
\cite{Dalfano99}.

In the present work, we avoid to use the Gross-Pitaevskii's theory and
we study the BEC in a phenomenological way  where the bose-gas
is considered as a many body system \cite{Ketterle}.
In particular, we study the ground state
of a system  of correlated bosonic atoms at zero  temperature,
trapped by a HO potential. The key quantity
for this effort is the two body density distribution (TBDD)
$ \rho({\bf r}_1,{\bf r}_2)$,
which expresses the joint probability
of finding two atoms at the positions ${\bf r}_1$ and ${\bf r}_2$,
respectively. In the mean field case, the TBDD is the product
$
\rho({\bf r}_1,{\bf r}_2)=\rho({\bf r}_1)\rho({\bf r}_2)
$
where $\rho({\bf r})$ is the density distribution (DD) of the system.
As the mean field approach fails to incorporate the
inter-particle interactions which are necessary for the description
of the correlated Bose system, we introduce the repulsive interactions
through the Jastrow correlation function \cite{Jastrow55}
$f(\mid{\bf r}_1-{\bf r}_2\mid)$. In this approximation the TBDD
has the form
$ \rho({\bf r}_1,{\bf r}_2)= N\rho({\bf r}_1)\rho({\bf r}_2)
f^2(r_{12}),$
 where the normalization constant $N$ ensures that
$\int\rho({\bf r}_1,{\bf r}_2)  d {\bf r}_1 d {\bf r}_2=1$
\cite{Fabrocini99,Moustakidis00,Moustakidis01}.

The calculations of the ground state properties of the correlated Bose system
(DD, rms radii, momentum distribution (MD) and kinetic energy (KE)) are made
using  four different expressions for the correlation function $f(r_{12})$.
Two of them are Jastrow type Gaussian functions similar to those
which have been used extensively in Nuclear Physics
\cite{Moustakidis00,Moustakidis01}, while, three of them are of hard sphere
form and they are more realistic in describing atomic and molecular
systems \cite{Brueckner57,Huang57,Wu59}.
The various expressions of the correlation function have one or two
parameters (the correlation parameter $\beta$ and the hard sphere radius
$r_0$).
In the case of the Jastrow type Gaussian function, analytical
expression of the one-body density matrix (OBDM) is derived
and is also used for the calculation of the corresponding
natural orbitals (NO's) and natural occupation numbers (NON's).
Finally, we give analytical or approximate expressions for the rms radius
and the mean kinetic energy $\langle T \rangle $ (in one case)
which  depend on the parameters $\beta$ or/and $r_0$.

The plan of the paper is as follows. In Sec. II the general definitions
related to the Bose system are considered. Details of the lowest-order
cluster expansion and analytical expressions of the DD (and in one
case of the MD) are given in Sec. III. Numerical results are reported
and discussed in Sec. IV,  while the summary of the present work
is given in Sec. V.

\section{General definitions}

Let $\Psi({\bf r}_1,{\bf r}_2,\cdots,{\bf r}_A)$ be the wave function
which describes the inhomogeneous atomic system. In the case where
this system is composed of bosonic atoms at zero temperature, all atoms
occupy the same single-particle ground state. The many body
ground state wave function is then a product of $A$ identical single
particle ground state wave functions. This ground state wave
function is therefore called the condensate wave function or
macroscopic wave function and has the form \cite{Ketterle99}
\begin{equation}
\Psi({\bf r}_1,{\bf r}_2,\cdots,{\bf r}_A)=
\psi_o({\bf r}_1)\psi_o({\bf r}_2) \cdots \psi_o({\bf r}_A) ,
\label{WF-1}
\end{equation}
where $\psi_o({\bf r})$ is the normalized to 1 ground state single
particle wave function describing a bosonic atom.
It is worth to indicate that Eq. (\ref{WF-1}) is valid even  when
 weak interactions are included. In this case the wave function
$\Psi({\bf r}_1,{\bf r}_2,\cdots,{\bf r}_A)$ is still, to a very good
approximation, a product of $A$ single particle wave functions which are
now obtained from the solution of a non-linear
Schr\"odinger equation, the well known Gross-Pitaevskii equation.

The density distribution $\rho({\bf r})$, which can be directly
observed in a non-destructive way, is defined as \cite{Ketterle99}
\begin{equation}
\rho({\bf r})=\int
\Psi^{*}({\bf r},{\bf r}_2,\cdots,{\bf r}_A)
\Psi({\bf r},{\bf r}_2,\cdots,{\bf r}_A)
 d  {\bf r}_2
 d  {\bf r}_3 \cdots  d  {\bf r}_A .
\label{OBDD-1}
\end{equation}
In the condensation case, where the wave function of the system is given
by Eq. (\ref{WF-1}), the DD, using  relation (\ref{OBDD-1}), takes
the form
\begin{equation}
\rho({\bf r})=\mid\psi_o({\bf r})\mid^2 .
\end{equation}

The TBDD $\rho({\bf r}_1,{\bf r}_2)$,
which is a key quantity in this work, is defined as
\begin{equation}
\rho({\bf r}_1,{\bf r}_2)=\int
\Psi^{*}({\bf r}_1,{\bf r}_2,\cdots,{\bf r}_A)
\Psi({\bf r}_1,{\bf r}_2,\cdots,{\bf r}_A)
 d  {\bf r}_3 \cdots  d  {\bf r}_A ,
\end{equation}
which in the condensation case, takes the simple form
\begin{equation}
\rho({\bf r}_1,{\bf r}_2)=\mid\psi_o({\bf r}_1)\mid^2
\mid\psi_o({\bf r}_2)\mid^2=\rho({\bf r}_1)\rho({\bf r}_2) .
\end{equation}
The TBDD is needed for the evaluation of the two-body properties of the
atomic system, mainly for the interaction energy. It gives also direct
information about the
correlations between the atoms of the system. This is the reason that
the calculation of this quantity is very significant for the study
of atomic or in general quantum many body systems.

Another quantity characterizing the atomic system is the
OBDM $\rho({\bf r}_1,{\bf r}_1')$
\cite{Naraschewski99,Lowdin55,Onsager56,Yang62,Pethick00}
which is defined as
\begin{equation}
\rho({\bf r}_1,{\bf r}_1')=\int
\Psi^{*}({\bf r}_1,{\bf r}_2,\cdots,{\bf r}_A)
\Psi({\bf r}_1',{\bf r}_2,\cdots,{\bf r}_A)
  d  {\bf r}_2  d  {\bf r}_3 \cdots  d  {\bf r}_A .
\end{equation}
The knowledge of the OBDM is also very important because
this quantity is connected to the position and to the momentum space.
The diagonal part of the OBDM gives the DD
$\rho({\bf r, r'})=\rho({\bf r})$, while the MD $n({\bf k })$
\cite{Baym96,Minguzzi00,Pitaevski99} is given
by a particular Fourier transform of it
\begin{equation}
n({\bf k})=\frac{1}{(2\pi)^3} \int \rho({\bf r}_1,{\bf r}_1')
\exp \left[i{\bf k}({\bf r}_1-{\bf r}_1')\right]
 d  {\bf r}_1   d  {\bf r}_1' .
\label{mom}
\end{equation}

In addition, the OBDM may be expanded in terms of its eigenfunctions
$\psi_i({\bf r})$ corresponding to the eigenvalues $n_i$, that is
\begin{equation}
\rho({\bf r}_1,{\bf r}_1')=
\sum_i n_i\psi_{i}^{*}({\bf r}_1)\psi_i({\bf r}_1') ,
\label{OBDM-NOFORM}
\end{equation}
where
\begin{equation}
\sum_{i}n_i=1 \ .
\end{equation}
The eigenfunctions $\psi_i({\bf r}),$ which are called natural
orbitals, and the eigenvalues $n_i$, which are called natural
occupation numbers, are obtained by diagonalizing the OBDM through
the eigenvalue equation
\begin{equation}
\int \rho({\bf r}_1,{\bf r}_1') \psi_i({\bf r}_1')  d {\bf r}_1'
=n_i\psi_i({\bf r}_1) .
\label{diag-rho}
\end{equation}
The condition, generally adopted, for the existence of condensation
is that there should be one eigenvalue $n_i$ which is of the order of
the number of the particles in the trap.

\subsection{Harmonic oscillator trap and Gross-Pitaevskii equation}

The first step for the study of the BEC at zero temperature is to
neglect the atom-atom interaction. In this case (named
independent particle model (IPM)), we consider that the atoms are confined
in an isotropic HO well and the Schr\"odinger equation takes the form
\begin{equation}
\left[-\frac{\hbar^2}{2m} \nabla^2 +\frac{1}{2}m\omega^2r^2\right]
\psi({\bf r})=\varepsilon \psi({\bf r}),
\end{equation}
where the ground state single-particle  wave function
$\psi({\bf r})$  has the form
\begin{equation}
\psi_{o}(r)=\left(\frac{1}{\pi b^2}\right)^{3/4}
\exp\left[-\frac{r^2}{2 b^2}\right],
\end{equation}
and the width $b$ is
\[
b=\left(\frac{\hbar}{m\omega}\right)^{1/2} .
\]
The normalization of the wave function is
$\int \mid \psi({\bf r}) \mid^2  d {\bf r}=1$,
while the DD has the form
$ \rho({\bf r})=\mid \psi_0({\bf r}) \mid^2 $.

It is obvious that for a system of non-interacting bosons in a HO
trap the condensate has the Gaussian form  of average width $b$.
If the atoms are interacting, the shape of the condensate can be changed
significantly with respect to the Gaussian \cite{Dalfano99}.
The ground-state properties of the condensate, for weakly interacting atoms,
are explained quite successfully by the non-linear equation,
known as Gross-Pitaevskii equation, of the form
\begin{equation}
\left[-\frac{\hbar^2}{2m} \nabla^2 + \frac{1}{2}m\omega^2 r^2+
A\frac{4\pi \hbar^2 a }{m} \mid \psi({\bf r}) \mid ^2\right]
\psi({\bf r})=\mu \psi({\bf r}) ,
\end{equation}
where $A$ is the number of the atoms, $a$ is the scattering length
of the interaction and $\mu$ is the chemical potential
\cite{Dalfano99}.
This equation has the form of a non-linear stationary Schr\"odinger equation,
and it has been solved for several types of traps using different
numerical methods \cite{Edwards95,Dalfano96c,Capuzzi,Cerboneschi}.
The presence of the third term, which is linear in
$A$ is responsible for the dependence of the
gas parameter $\chi$  on the density of the system.
That dependence is negligible in the case of the IPM.
In the phenomenological approach which we consider in this work,
there is not a direct dependence  between the condensation and the number
of the atoms.
The inter-particle correlations is incorporated in the mean field
only by the correlation function which, in a way, depends
on the effective size of the atoms.

\section{Lowest-order cluster expansion}

A dilute BE atomic system can be studied using the lowest-order approximation
(LWOA)\cite{Fabrocini99}.
In the LWOA the two-body density matrix (TBDM) has the form
\begin{equation}
\rho({\bf r}_1,{\bf r}_2;{\bf r}_1',{\bf r}_2)=
N \rho({\bf r}_1,{\bf r}_1' ) \rho({\bf r}_2)
f(\mid{\bf r}_1' - {\bf r}_2\mid)
f(\mid{\bf r}_1-{\bf r}_2\mid),
\label{TBDM-1}
\end{equation}
where $f({\bf r}_1,{\bf r}_2 )$ is the Jastrow correlation function, which
depends on the inter-particle distance and $N$ is the normalization factor.

The diagonal part of $\rho({\bf r}_1,{\bf r}_2;{\bf r}_1',{\bf r}_2)$
is the TBDD
\begin{equation}
\rho({\bf r}_1,{\bf r}_2)= N \rho({\bf r}_1)\rho({\bf r}_2)
f^2(r_{12}) ,
\label{TBDD-1}
\end{equation}
while the OBDM is
\begin{equation}
\rho({\bf r}_1,{\bf r}_1')= \int
\rho({\bf r}_1,{\bf r}_2;{\bf r}_1',{\bf r}_2) d {\bf r}_2 .
\label{OBDM-1}
\end{equation}

The DD, which is the diagonal part of $\rho({\bf r}_1,{\bf r'}_1)$,
can also be obtained from the integral
\begin{equation}
\rho({\bf r})=N \int\rho({\bf r})\rho({\bf r}_2)
f^2(\mid{\bf r}-{\bf r}_2\mid)  d  {\bf r}_2
\label{cor-dd1}
\end{equation}

In the mean field case, using Eq. (\ref{OBDM-NOFORM}),
the OBDM becomes
\begin{equation}
\rho({\bf r}_1,{\bf r}_1')=\psi_{0}^{*}({\bf r}_1) \psi_0({\bf
r}_1') ,
\end{equation}
where $\psi_0({\bf r})$ is the uncorrelated ground state wave
function. In the case of the inclusion of the inter-particle
interactions between the atoms, which give rise to the depletion
of the condensate, the OBDM takes the form \cite{Stringari01}
\begin{equation}
\rho({\bf r}_1,{\bf r}_1')= n_0 \psi_{0}^{*}({\bf r}_1)
\psi_{0}({\bf r}_1')+ \sum_{i\neq 0}n_i \psi_{i}^{*}({\bf
r}_1) \psi_{i}({\bf r}_1') ,
\end{equation}
where the sum $\sum_{i\neq 0}n_i \psi_{i}^{*}({\bf r}_1)
\psi_{i}({\bf r}_1')$ is the contribution arising from the
atoms out of the condensate.

The NO's $\psi_{i}({\bf r}_1)$ and the NON's
$n_i$ are obtained by diagonalizing the
OBDM through the eigenvalue Eq. (\ref{diag-rho}).
This is made expanding the OBDM in Legendre polynomials
\begin{equation}
\rho({\bf r},{\bf r}')=\rho(r, r',\cos\omega_{rr'})=
\sum_{l=0}^{\infty}\rho_l(r,r') P_l(\cos\omega_{rr'}) ,
\label{NO-1}
\end{equation}
where
\begin{equation}
\rho_l(r,r')=\frac{2l+1}{2}\int_{-1}^{1}
\rho(r, r',\cos\omega_{rr'}) \
P_l(\cos\omega_{rr'}) \ d (\cos\omega_{rr'}) ,
\label{NO-2}
\end{equation}
and substituting Eq.(\ref{NO-1}) into Eq. (\ref{diag-rho}). Using
the addition theorem of the spherical harmonics and integrating
over the angle $\Omega_{r_{1}'}$, the eigenvalue problem takes the form
\begin{equation}
4\pi \int_{0}^{\infty}\rho_{l}(r,r')\varphi_{nl}^{NO}(r')
{r'}^2{\rm d} r'=n_{nl}^{NO}\varphi_{nl}^{NO}(r) ,
\label{NO-3}
\end{equation}
where $\varphi_{nl}^{NO}(r)$ is the radial part of  $\psi_{i}({\bf r})$
\[
\psi_{i}({\bf r})=\varphi_{nl}^{NO}(r)Y_{lm}(\Omega_r)
\ . \]

In the present work, we use four different expressions of the correlation
function $f(r_{12})$ for the study of the BE atomic systems. These
expressions are
\begin{eqnarray}
\label{case-1}
{\rm Case \ 1}\qquad & &f(r)=1-\exp\left[-\beta r^{2}\right],\\
& & \nonumber\\
\label{case-2}
{\rm Case\ 2} \qquad & &
f(r)=\left\{ \begin{array}{cc}
\left(1-r_o/r \right)^{1/2}, &  \mbox{$r>r_0$} \\
                              0, &  \mbox{$r<r_0$}
                              \end{array}
                       \right.  ,    \\
& & \nonumber\\
\label{case-3}
{\rm Case\ 3}\qquad & &
f(r)=\left\{ \begin{array}{cc}
1-\exp\left[-\beta(r^2-r_{0}^{2})\right], &  \mbox{$r>r_0$} \\
                              0, &  \mbox{$r<r_0$}
                              \end{array}
                       \right. ,  \\
& & \nonumber\\
\label{case-4}
{\rm Case\ 4} \qquad & &
f(r)=\left\{ \begin{array}{cc}
\left(1-\frac{\exp\left[-\beta(r^2-r_{0}^{2})\right]}{r/r_0} \right)^{1/2}, &
\mbox{$r>r_0$} \\
                              0, &  \mbox{$r<r_0$}
                              \end{array}
                       \right.   ,
\end{eqnarray}
where $r=\mid {\bf r}_1-{\bf r}_2 \mid$.

The correlation function $f(r)$ in case 1 for large values of $r$ goes to 1
and it goes to 0 for $r \rightarrow 0$. It is obvious that the effect of
correlations introduced by the function $f(r)$ becomes large when the
correlation parameter $\beta$ becomes small and vice versa.
Case 2 is the classical hard sphere correlation function, which has been
used extensively in atomic and molecular physics and depends on the
hard sphere radius $r_0$.
Case 3, which is a generalization of case 1, depends also on the "hard
sphere" radius $r_0$ of the interaction between the atoms.
In the limit $r_0 \rightarrow 0$, $f(r)$ goes to that of case 1.
Finally, case 4 is a generalization of case 2
including a Gaussian function with the additional parameter $\beta$.
In the limit $\beta \rightarrow 0$ $f(r)$ goes to that of case 2.

The above defined correlation functions were used to find analytical
expressions of the DD through Eqs. (\ref{OBDM-1}) or/and (\ref{cor-dd1}).
In case 1 the OBDM and the MD was calculated also analytically through Eq.
(\ref{OBDM-1})  and (\ref{mom}) respectivelly, while the NO's and the
NON's are calculated through Eq. (\ref{NO-3}).

\subsection{Analytical expressions in case 1}
In case 1 we found the analytical expressions of the DD and MD from
the OBDM which have been calculated from Eq. (\ref{OBDM-1}) and has the form
\begin{equation}
\rho({\bf r},{\bf r}')=
\frac{N}{\pi^{3/2}b^3}\left[O_1({\bf r},{\bf r}')-O_{21}({\bf r},{\bf r}')-
O_{22}({\bf r},{\bf r}')+O_{23}({\bf r},{\bf r}')\right] ,
\label{cluster-11}
\end{equation}
where $N$ is the normalization factor of the form
\begin{equation}
N=\left[1-\frac{2}{(1+2y)^{3/2}}+\frac{1}{(1+4y)^{3/2}}\right]^{-1} ,
\label{norm-1}
\end{equation}
and the one- and the two-body terms of the cluster expansion have
the forms
\begin{eqnarray}
\label{case1-o1r}
O_1({\bf r},{\bf r}')&=&
\exp\left[-\frac{r_b^2+{r_b'}^2}{2}\right], \\
& & \nonumber\\
O_{21}({\bf r},{\bf r}')&=& \frac{1}{(1+y)^{3/2}}
\exp\left[-\frac{1+3y}{1+y}\frac{r_b^2}{2}-\frac{{r_b'}^2}{2}\right] , \\
& & \nonumber\\
O_{22}({\bf r},{\bf r}')&=&O_{21}({\bf r}',{\bf r}), \\
& & \nonumber\\
O_{23}({\bf r},{\bf r}')&=& \frac{1}{(1+2y)^{3/2}}
\exp\left[-(1+2y)\frac{r_b^2+{r_b'}^2}{2}\right]
\exp\left[\frac{y^2}{1+2y}({\bf r}_b+{\bf r}_b')^2\right] ,
\end{eqnarray}
where $r_b=r/b$ and $y=\beta b^2$.

The analytical expression of the DD can be found from Eq.
(\ref{cluster-11}), putting $r'=r$, while the MD can be found analytically
using Eq. (\ref{mom}). It takes the form
\begin{equation}
n(k)= \frac{N b^3}{\pi^{3/2}} \left[
\exp\left[-k_b^2\right]
- \frac{2}{(1+3y)^{3/2}}
\exp\left[-\frac{1+2y}{1+3y}k_b^2\right]
+
\frac{1}{(1+2y)^{3/2}(1+4y)^{3/2}}
\exp\left[-\frac{1}{1+2y}k_b^2\right]    \right] ,
 \label{cluster-nk}
\end{equation}
where $k_b =kb$.

The expressions of $\rho(r)$ and $n(k)$, given by Eq.
(\ref{cluster-11}) (for $r'=r$) and (\ref{cluster-nk}), respectively,
have been used to find the analytical expressions of the mean square
radius and kinetic energy of the trapped gas, through the integrals
\begin{eqnarray}
\label{rad-ms}
\langle r^2 \rangle & =& 4\pi \int_0^{\infty} \rho(r) r^4 d r ,\\
\nonumber\\
\label{kinetic}
\langle T \rangle & =& \frac{\hbar^2}{2m} 4\pi \int_0^{\infty} n(k) k^4 d k,
\end{eqnarray}
respectivelly. The expressions we found, for $\langle r^2 \rangle$
and $\langle T \rangle$, are
\begin{eqnarray}
\label{rad-1}
\langle r^2 \rangle& =&Nb^2 \left[ \frac{3}{2} -
3\frac{1+y}{(1+2y)^{5/2}}
+\frac{3}{2}\frac{1+2y}{(1+4y)^{5/2}} \right], \\
& & \nonumber\\
\label{kinetic-1}
\langle T \rangle& =& N \hbar\omega  \left[ \frac{3}{4} -
\frac{3}{2}\frac{1+3y}{(1+2y)^{5/2}}
+ \frac{3}{4}\frac{1+2y}{(1+4y)^{3/2}}\right] .
\end{eqnarray}

These expressions, which for a given HO trap are functions of the correlation
parameter $y$, could be used to define $y$ from Eq. (\ref{rad-1}),
if the rms radius of the trapped atoms is known and then to define
the $\langle T \rangle$ from Eq (\ref{kinetic-1}) and vice versa.
For very large values of $y$ Eqs. (\ref{rad-1}) and (\ref{kinetic-1})
give the HO expressions of $\langle r^2 \rangle$ and $\langle T \rangle$, i.e.
$\langle r^2 \rangle = \frac{3}{2} b^2$ and
$\langle T \rangle = \frac{3}{4}\hbar\omega$, respectively.

\subsection{Analytical expressions in case 2}
In case 2, where the correlation function is given by Eq.
(\ref{case-2}), the DD defined by Eq. (\ref{cor-dd1}), is written
\begin{equation}
\rho(r)= \frac{N}{b^3 \pi^{3/2}}
\left[ O_1(r) - \frac{r_{0b}}{2r_b} \exp \left[-r_b^2 \right]
\left(\frac{}{} {\rm Erf}(r_b +r_{0b}) - {\rm Erf}(-r_b + r_{0b})
\right) \right] ,
\label{cluster-rho2}
\end{equation}
where $N$ is the normalization factor of the form
\begin{equation}
N=\left[1-{\rm Erf}\left( \frac{r_{0b}}{\sqrt{2}} \right) \right]^{-1} ,
\label{case2-norm}
\end{equation}
and
\[
{\rm Erf}(z)= \frac{2}{\sqrt{\pi}} \int_0^z {\rm e}^{-t^2} dt  .
\]
The one-body term $O_1(r)$ has the form
\begin{equation}
O_1(r)=\exp \left[-r_b^2 \right]
\left[ 1 - \frac{1}{2r_b \sqrt{\pi}} \left(
\exp\left[-(r_b+r_{0b})^2\right]
- \exp\left[-(r_b-r_{0b})^2\right]  \right)
- \frac{1}{2}\left( {\rm Erf}(r_b + r_{0b})
+ {\rm Erf}(-r_b + r_{0b}) \right) \right] .
\label{case2-o1}
\end{equation}

\subsection{Analytical expressions in case 3}
In case 3, where the correlation function is given by Eq.
(\ref{case-3}), the DD defined by Eq. (\ref{cor-dd1}), is written
\begin{equation}
\rho(r)= \frac{N}{b^3 \pi^{3/2}}
\left[O_1(r)-2O_2(r,\beta)+O_2(r,2\beta)\right],
\label{cluster-rho3}
\end{equation}
where $N$ is the normalization factor of the form
\begin{equation}
N = \left[ N_1 - 2 N_2(\beta) + N_2(2 \beta)\right]^{-1} ,
\label{case3-norm}
\end{equation}
where
\begin{eqnarray}
\label{norm-N1}
N_1 &=& 1 - {\rm Erf}\left(\frac{r_{0b}}{\sqrt{2}}\right)+
\sqrt{\frac{2}{\pi}}r_{0b}\exp\left[-\frac{r_{0b}^{2}}{2}\right], \\
& &\nonumber\\
\label{norm-N2}
N_2(\beta)&=&
 \frac{\exp\left[ yr_{0b}^{2} \right]}
{(1+2y)^{3/2}} \left[ 1 -
{\rm Erf}\left( r_{0b}\sqrt{\frac{1+2y}{2}} \right) \right] +
\sqrt{ \frac{2}{\pi}}
\frac{r_{0b}}{1+2y} \exp\left[ -\frac{r_{0b}^{2}}{2} \right] ,
\end{eqnarray}
where $r_{0b}=r_0/b$. The factor $N_2(2\beta)$ can be found from the factor
$N_2(\beta)$ replacing $y \rightarrow 2y$.

The one-body term is the same as in case 2 and is given by Eq.
(\ref{case2-o1}), while the two-body term $O_2(r,\beta)$ has the form
%
%
\begin{eqnarray}
\label{case2-o2}
O_2(r,\beta)&=&
\frac{\exp \left[y r_{0b}^{2}\right]}{(1+y)^{3/2}}
\exp\left[-\frac{1+2y}{1+y}r_b^2  \right]
\left[1 -
\frac{1}{2}
\left( {\rm Erf}\left(\frac{r_b+r_{0b}(1+y)}{(1+y)^{1/2}}\right) +
{\rm Erf}\left(\frac{-r_b+r_{0b}(1+y)}{(1+y)^{1/2}}\right)
\right)  \right] \nonumber\\
\nonumber\\
&-&\frac{\exp[-r_{b}^2]}{2 r_b \sqrt{\pi} (1+y) }
\left[ \exp\left[-(r_b+r_{0b})^2\right] -\exp\left[-(r_b-r_{0b})^2\right]
  \right] .
\label{case3-o2}
\end{eqnarray}
The term $O_2(r,2\beta)$ can be found from the term $O_2(r,\beta)$
replacing $y \rightarrow 2y$.


\subsection{Analytical expressions in case 4}
In case 4, where the correlation function is given by Eq.
(\ref{case-4}), the DD defined by Eq. (\ref{cor-dd1}), is written
\begin{equation}
\rho(r)= \frac{N}{b^3\pi^{3/2}}
\left[ O_{1}(r) - O_2(r,\beta) \right],
\label{cluster-rho4}
\end{equation}
where $N$ is the normalization factor of the form
\begin{equation}
N=
\left[ 1-{\rm Erf}\left( r_{0b}/ \sqrt{2} \right) +
\frac{2^{3/2}}{\sqrt{\pi}}
\frac{y r_{0b}}{1+2y} \exp\left[- r_{0b}^2/ 2 \right]  \right]^{-1} .
\label{case4-norm}
\end{equation}

The term $O_1(r)$ is the same as in cases 2 and 3, while the two-body term
is of the form
\begin{eqnarray}
O_2(r,\beta)&=&\frac{r_{0b}}{2r_b}
\frac{\exp\left[ yr_{0b}^{2} \right]}{(1+y)^{1/2}}
\exp\left[ -\frac{1+2y}{1+y}r_b^2 \right]
\left[{\rm Erf}\left(\frac{r_b+r_{0b}(1+y)}{(1+y)^{1/2}}\right)-
{\rm Erf}\left( \frac{-r_b+r_{0b}(1+y)}{(1+y)^{1/2}} \right) \right].
\label{case4-02}
\end{eqnarray}

The above expressions of the DD (and of the MD in case 1) depend on two
parameters in cases 1 and 2 and on three parameters in cases 2 and 4, i.e.
the HO parameter $b$, the correlation parameter $\beta$ or/and the hard
sphere radius $r_0$.
We may note, however that their dependence on $\beta$ and $r_0$ is only
through the dimensionless quantities $y=\beta b^2$ and $r_{0b}=r_0/b$.

\section{Results and discussion}
The calculation of the DD of trapped Bose gas,
confined in an isotropic HO potential with length $b=10^4 \AA$,
has been carried out on the basis of Eqs. (\ref{cluster-11}),
(\ref{cluster-rho2}), (\ref{cluster-rho3}) and (\ref{cluster-rho4}) which
have been found using four different expressions for the correlation
function. The dependence of the DD on the parameters $y$ and $r_0$ are
shown in Figs. 1 to 3.

The DD, in case 1, for various values
of the parameter $y$, including also the uncorrelated case ($y=\infty$),
has been plotted in Fig. 1a.
It is seen that, the large values of $y$ ($y>10$) correspond to the
Gaussian distribution (HO case), while when $y$ becomes small enough ($y<1$)
the DD spreads out as in Gross-Pitaevskii's theory.
For $y>10$ the effect of correlations is small, while for very large
correlations ($y \lesssim 0.1$) the DD is modified entirely from the Gaussian
form which originates from the HO trap.
The same effect of the inter-particle correlations on the DD appears also in
case 2 where the correlation function is that of the hard sphere. This is
seen in Fig. 1b where the DD has been plotted for various
values of the hard sphere radius $r_0$. The effect of the correlations
becomes significant for $r_0 > 1000 \AA$.
Thus the inter-particle
correlations become large when $r_0$ approaches the characteristic
trap length $b$. It is noted that, in Gross-Pitaevskii's theory, the
difference between the correlated DD and the uncorrelated one
becomes significant for smaller values of the parameter $r_0$
\cite{Dalfano96a,DuBois 01}.

In Fig. 2 the DD in case 3 for three values of the parameter
$y$ and various values of the parameter $r_0$ is shown.
It is seen that small values of $y$ result to a depression of the
central part of the DD independent of the values of $r_0$. This is obvious
in Fig. 2, where for  $y=0.1$ and $r_0>500 \AA$, the maximum of the DD is
moved from the center ($r/b=0$) to $r/b \simeq 1$. The same behavior of the
DD in case 4 can be seen in Fig. 3, where it is plotted for the same values
of $y$ and $r_0$ as in Fig. 2.
The only difference is that the same depression of the
DD in the central region appears for smaller values of the parameter $y$.

The rms radius of the Bose gas has also been calculated analytically from
Eq. (\ref{rad-1}) in case 1 and numerically in the other cases from
Eq. (\ref{rad-ms})
for various values of the parameters $y$ or/and $r_0$. The results are shown
in Figs. 4 and 5.
From Fig. 4a, which corresponds to the DD calculated in case 1,
it is seen that  $\langle r_{b}^2\rangle ^{1/2}$ ($r_b=r/b$)
is a decreasing function of the
parameter $y$ and for $y > 10$ approaches the rms radius of the
uncorrelated case.
If we expand the expression of $\langle r_{b}^2\rangle ^{1/2}$,
given by Eq. (\ref{rad-1}), in powers of $y^{-1}$ and truncate the
expansion to the $y^{-3}$ power, the following approximate expression is
obtained,
\begin{equation}
\langle r_b^2\rangle = \frac{3}{2} + 0.4366 y^{-3/2}
- 0.60432 y^{-5/2} + 0.2541 y^{-3} .
\label{approx-rad1}
\end{equation}

The values of $\langle r_{b}^2\rangle ^{1/2}$ calculated from that expression are
very close to the ones calculated from Eq. (\ref{rad-1}) for $y>2$.

The behavior of $\langle r_{b}^2\rangle ^{1/2}$, in case 2, as function of the
parameter $r_0$ is shown in Fig 4b. In this case,
$\langle r_{b}^2\rangle ^{1/2}$ is an increasing function of $r_0$.
We fitted the form $\langle r_b^2\rangle = \frac{3}{2} + C r_{0b}^{3/2}$
to the values of the rms radius obtained in case 2 and we found the
expression
\begin{equation}
\langle r_b^2\rangle = \frac{3}{2} + 0.3978 r_{0b}^{3/2} .
\label{approx-rad2}
\end{equation}
This simple expression gives values for $\langle r_{b}^2\rangle ^{1/2}$ 
which are very close to the ones calculated from 
the DD as can be seen from Fig. 4b.

The dependence of $\langle r_{b}^2\rangle ^{1/2}$ on $y$ and $r_0$, in cases
3 and 4, are shown in Fig. 5, where it is plotted
versus $r_0$ for three values of the parameter $y$.
It is seen that, in both cases for the same value of $y$,
$\langle r_{b}^2\rangle ^{1/2}$ is an increasing function of $r_0$, as in case 2,
while for the same value of $r_0$ it is a decreasing function of $y$, as in
case 1. This behavior of $\langle r_{b}^2\rangle ^{1/2}$
as well as Eqs. (\ref{approx-rad1}) and (\ref{approx-rad2})
lead us to fit the numerical values of $\langle r_{b}^2\rangle ^{1/2}$
in cases 3 and 4 with two functions, which are combinations of the expressions
of $\langle r_{b}^2\rangle ^{1/2}$ of cases 1 and 2. 
For the two cases we found the expressions
\begin{eqnarray}
\label{approx-3}
{\rm Case \ 3}\qquad& & \langle r_b^2\rangle = \frac{3}{2}
+ \left( 0.2916 y^{-3/2} -0.1290 y^{-5/2} +0.0323 y^{-3} \right) +
0.1327 r_{0b}^{3/2}, \\
& & \nonumber\\
\label{approx-4}
{\rm Case \ 4}\qquad& & \langle r_b^2\rangle = \frac{3}{2}
+ \left( 5.4556 y^{-3/2} -7.3299 y^{-5/2} + 2.1458 y^{-3}\right) r_{0b}^{3/2} .
\end{eqnarray}
From Fig. 5, it is seen that the numerical values of
$\langle r_{b}^2\rangle ^{1/2}$ calculated in cases 3 and 4 from the DD
are very close to the ones calculated from Eqs. (\ref{approx-3}) and
(\ref{approx-4}), respectively.

The NO's and the NON's have been calculated, in case 1,
by diagonalizing the OBDM through Eq. (\ref{NO-3}).
The NON $n_{1s}$, gives directly
the condensation fraction $n_0$  as a result of the repulsive
interaction between the atoms of the Bose gas at zero temperature.
The NON's for the ground state and for some excited states are given
in Table 1. It seems that, for strong correlations, a fraction
of atoms spread out into many higher excited states. The condensation fraction
$n_0$, versus the parameter $1/y$ is plotted in Fig. 6. 
From that figure and from Table 1 it is seen that the
effect of the correlations on $n_0$ is small
and all the atoms occupy the ground state, when $y>10$. The effect
of the correlations is prominent when $y<10$, while the decrease of the
parameter $y$ (large correlations) induces a significant depletion
of the condensated atoms exciting them into higher states.

The NO's
of the states $1s$, $1p$ and $1d$ for $y=1$ are shown in Fig. 7.
It is seen that the interatomic correlations
in the $1s$-state 
spread out the ground state wave function and consequently
the condensation is appears in the outer region of the trap.
From the same figure it is obvious that the NO's of 
excited states are much more localized in coordinate space than the
equivalent HO orbitals.

The MD in case 1 can be
calculated analytically from Eq. (\ref{cluster-nk}) 
or by Fourier transform of the NO's.
The MD calculated analytically 
for various values of the parameter $y$ has been plotted
in Fig. 8a.  It is seen that the large values of $y$ ($y>10$)
correspond to the Gaussian distribution, while when $y$ becomes small enough
($y<1$) the MD has a sharp maximum for $k=0$.
The MD of the $1s$-state NO as well as of the  residual excited states
for $y=0.1$ are shown and compared with the total MD in Fig. 8b.
It is obvious that
although the $1s$-state NO gives the main contribution
to the MD, the NO's of the excited state contribute to the MD mainly
in the large values of the momentum $k$. 

The dependence of the mean kinetic energy $\langle T \rangle$
on the parameters $y$  calculated analytically in case 1,
using Eq. (\ref{kinetic-1}), is presented in Fig. 9. It is seen that 
$\langle T \rangle$ has a maximum for $y\simeq2.5$. It is interesting
to note that for the same value of the parameter $y$ the NON's of the
states $1d$  and $1f$ have a maximum as can be seen from Table 1.
The contribution of the $1s$-state NO and of all the excited states to  
$\langle T \rangle$ are shown in the same figure. It is seen that,
for large values of the parameter
$y$ (weak correlations) the main contribution to 
$\langle T \rangle$ comes from the $1s$-state  NO, while
for large correlations there is a significant contribution
coming from the NO's of the excited states.

An approximate expression of  $\langle T \rangle$, similar to the one
of $\langle r_b^2\rangle$ can be found 
if we expand
the expression of $\langle T \rangle$, given by Eq. (\ref{kinetic-1}), in
powers of $y^{-1}$ and truncate the expansion to the $y^{-2}$ power.
The approximate expression has the form
\begin{equation}
\langle T \rangle /\hbar \omega = \frac{3}{4} + 0.1875 y^{-1/2}
-0.3355 y^{-3/2} + 0.1092 y^{-2} .
\label{approx-KE1}
\end{equation}
The values of $\langle T \rangle$ calculated from that expression are
very close to the ones calculated from Eq. (\ref{kinetic-1}) for $y > 2.5$.

A final comment is appropriate.  The Gross-Pitaevskii equation
explains successfully the ground-state properties of the condensate for
weakly interacting atoms, i.e. for small correlations between the atoms.
The method of the present work can be used for larger correlations,
provided that the expansions of the TBDM and the OBDM contain higher order
terms.
It should be noted also that, in the present work there is not a direct
dependence between the condensation and the number of the atoms.
The inter-particle correlations are incorporated in  the mean field only
by the correlation function which, in some way, depends on the effective
size of the atoms.
However, the condensate fraction can be found by calculating the
natural orbitals and the natural occupation numbers by diagonalizing
the OBDM from Eq. (\ref{diag-rho}). These calculations have been made in
case 1, while it is a very difficult task in the other cases.

\section{Summary}

The effect of the inter-particle correlations between Bose atoms
at zero temperature is examined using a phenomenological way to
incorporate the atomic correlations.
That is made by introducing the Jastrow correlation function in the TBDD.
We examined the effect of correlations using four different correlation
functions.
The introduction of correlations change the shape of the DD and MD
comparing with the Gaussian form, which corresponds to IPM.
There is a reduction of the DD in the central region of the atomic system,
and so an increase of the rms radius of the system, while the MD,
which is calculated in case 1,
increases in the  region of small $k$ and so there is a decrease of
the mean kinetic energy of the system. In case 1 where the mean
kinetic energy is calculated also analytically, there is a maximum for 
a certain value of the correlation parameter.
Finally in case 1 the natural orbitals  and the natural
occupation numbers have been calculated  and consequently the condensate
fraction has been obtained for different values of the parameter $y$.

\acknowledgments
The authors would like to thank Dr. C.P. Panos for useful comments on the
manuscript.


\begin{table}
\caption{The values of the natural occupation numbers in case 1
for different values of the correlation parameter $y$ .}
\begin{center}
\begin{tabular}{l l c c c c c c c c }
 & & & & & &   \\
$y$ & $n_{1s}$ & $n_{1p}$ & $n_{1d}$  & $n_{1f}$ & Sum
 \\
\hline
 & & & & & &  \\
100  & 0.99988     &   -       &    -          &  -         & 0.99988 \\
10   & 0.99634     & 0.00063   & 0.00042       &0.00042     & 0.99781  \\
5    & 0.99055     & 0.00273   & 0.00108       &0.00108     & 0.99544   \\
2.5  & 0.97771     & 0.00960   & 0.00186       &0.00186     & 0.99103   \\
1    & 0.94422     & 0.03462   & 0.00172       &0.00172     & 0.98228   \\
0.5  & 0.90815     & 0.06830   & 0.00082       &0.00082     & 0.97809   \\
0.1  & 0.83097     & 0.15185   & 0.00001       & 0.00001    & 0.98284  \\
0.01 & 0.79273     & 0.19414   &   -           & -          & 0.98687
\end{tabular}
\end{center}
\end{table}

\newpage
\begin{figure}
\begin{center}
\begin{tabular}{lr}
{\epsfig{figure=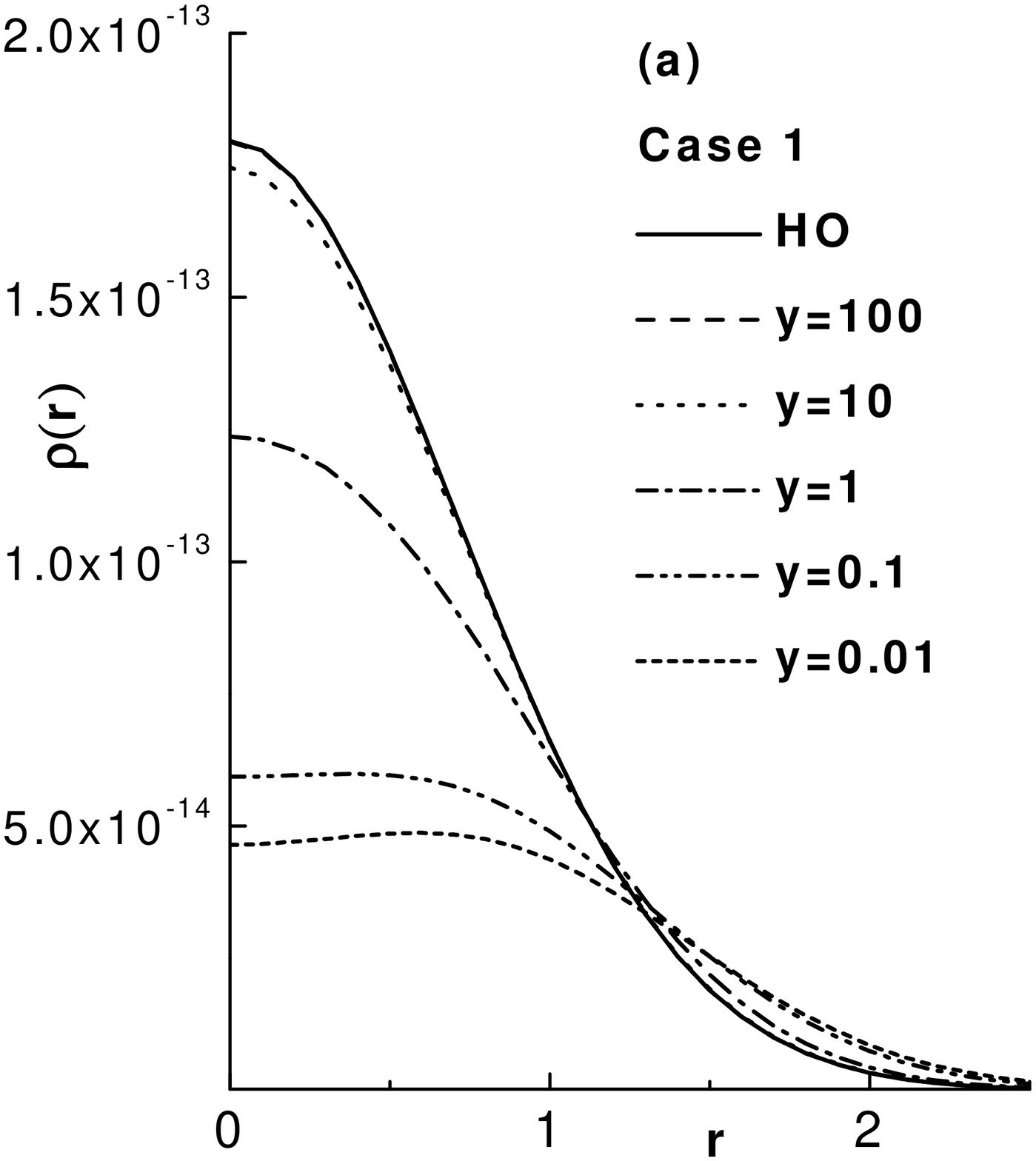,width=5cm} }&
{\epsfig{figure=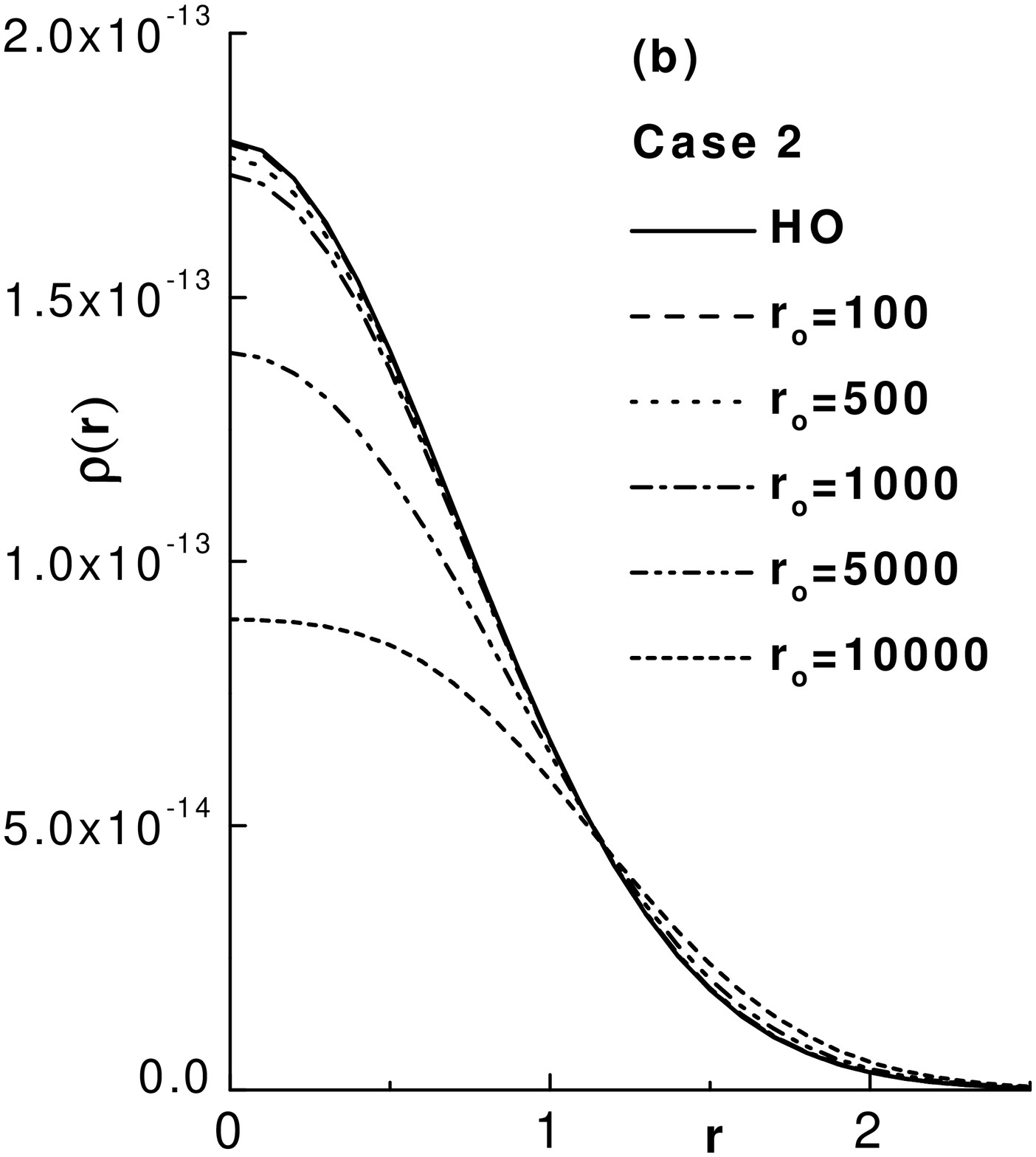,width=5cm} }\\
\end{tabular}
\end{center}
\caption{The DD $\rho(r)$ in cases 1 and 2 versus $r$ for various
values of the parameters $y$ and $r_0$, respectively. $r$ is in units of the
trap lenght $b$, $\rho(r)$ in $\AA^{-3}$ and $r_0$ in $\AA$.
The normalization is $\int \rho({\bf r})  d  {\bf r}=1$.}
\end{figure}

\begin{figure}
\begin{center}
\begin{tabular}{ccc}
{\epsfig{figure=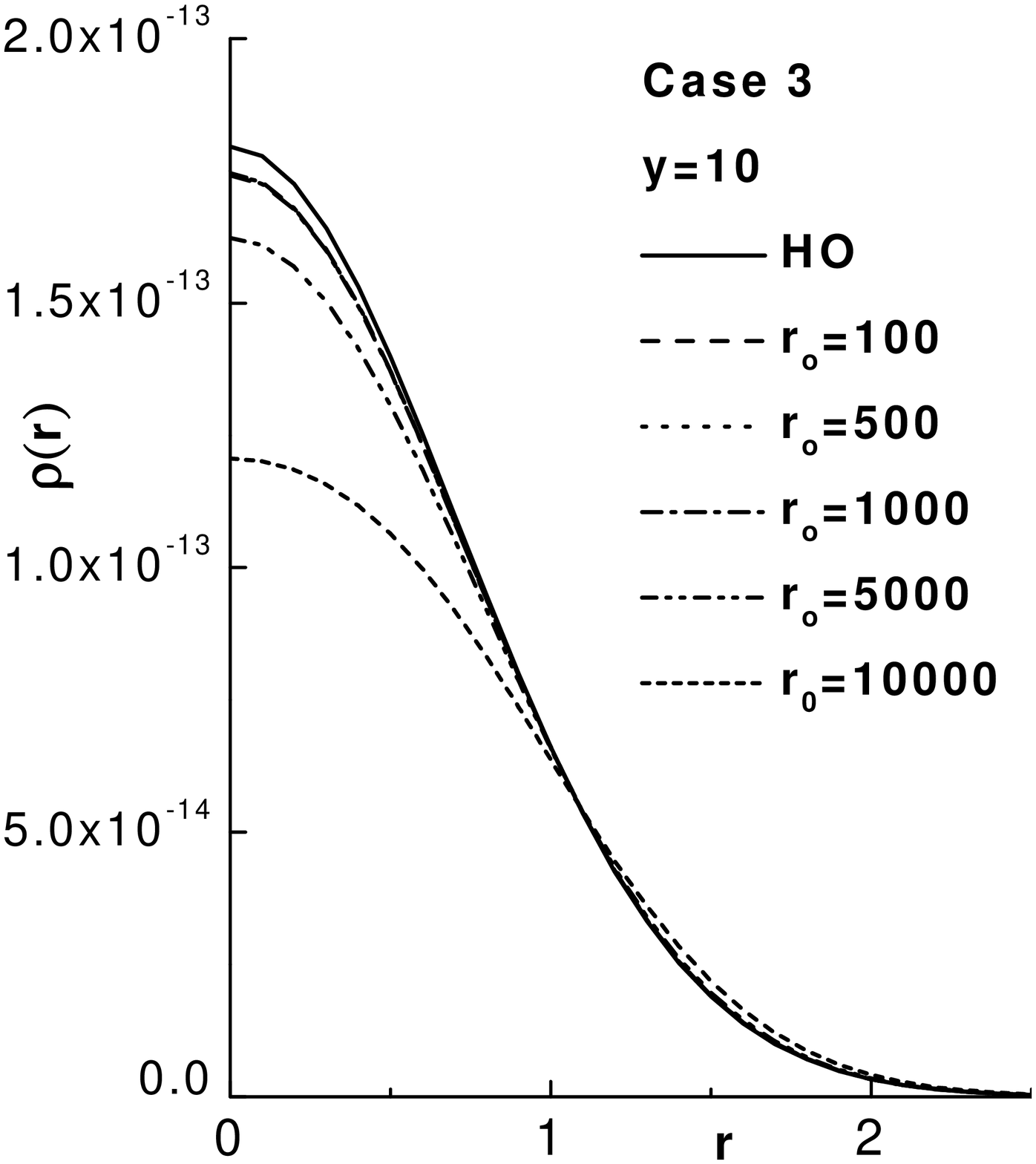,width=5.cm} }&
{\epsfig{figure=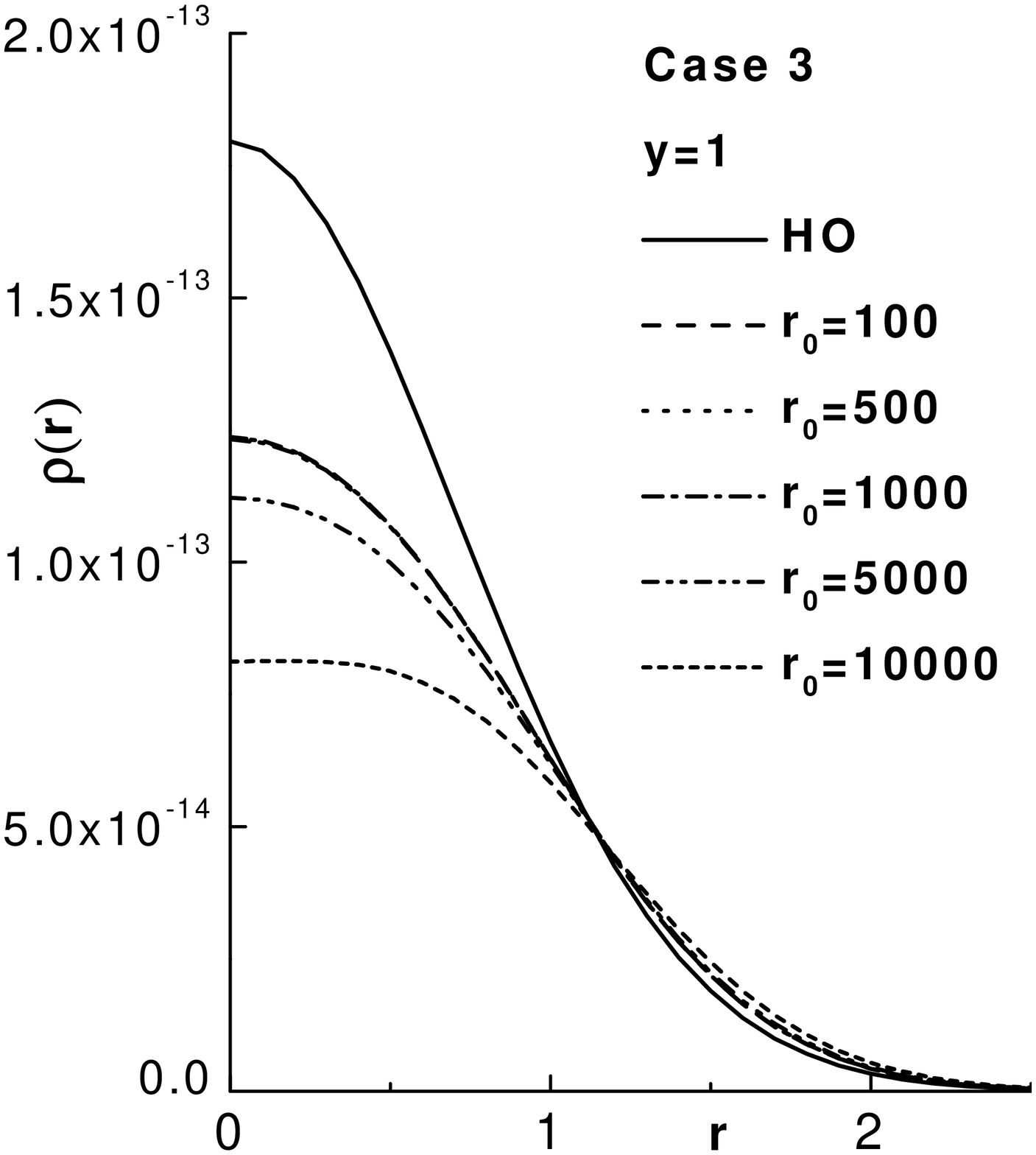,width=5.cm} }&
{\epsfig{figure=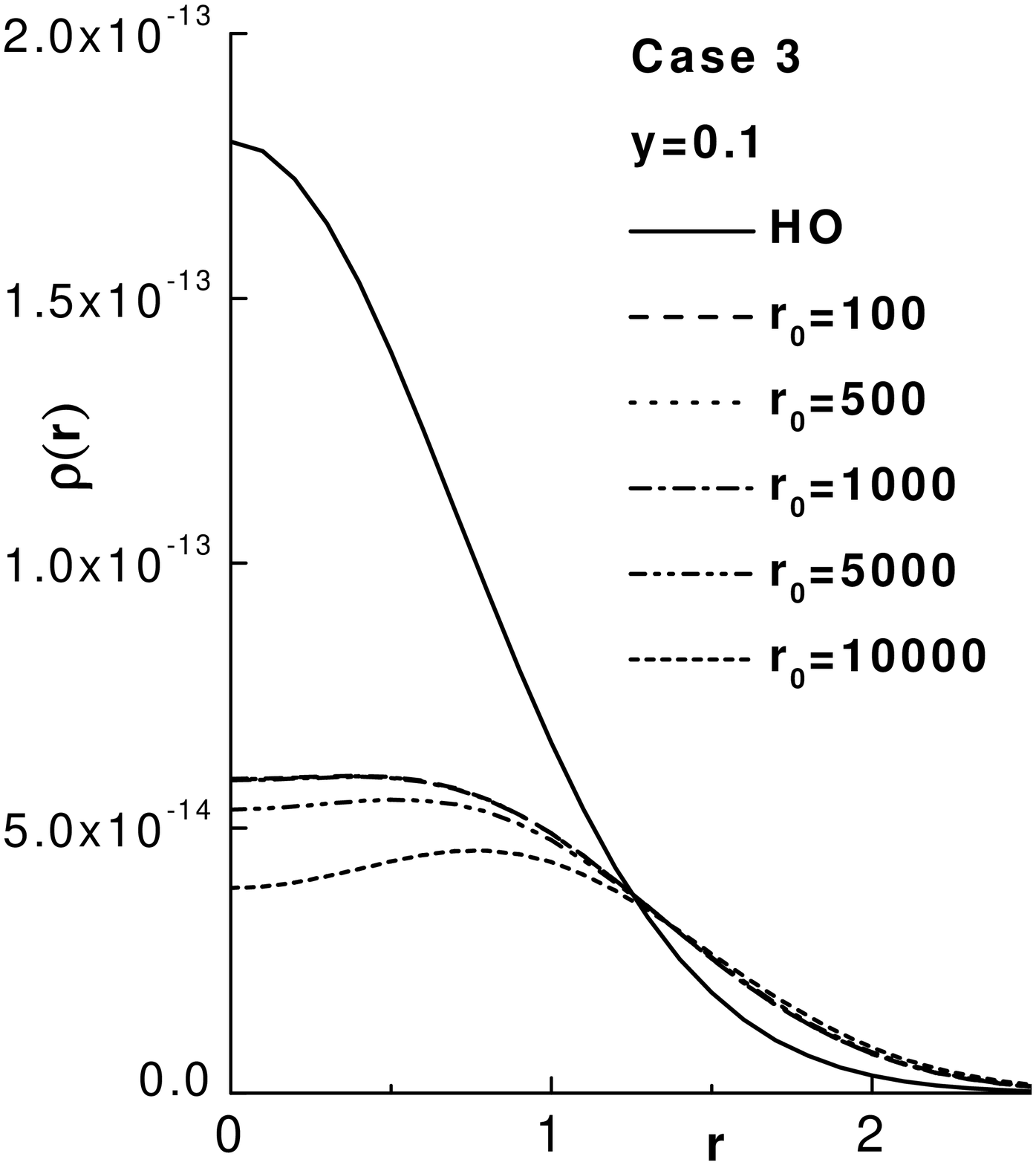,width=5.cm}}
\end{tabular}
\end{center}
\caption{The DD in case 3 versus $r$ for various values of the parameters
$y$ and $r_0$. The units and the normalization are as in Fig. 1.}
\end{figure}

\begin{figure}
\begin{center}
\begin{tabular}{ccc}
{\epsfig{figure=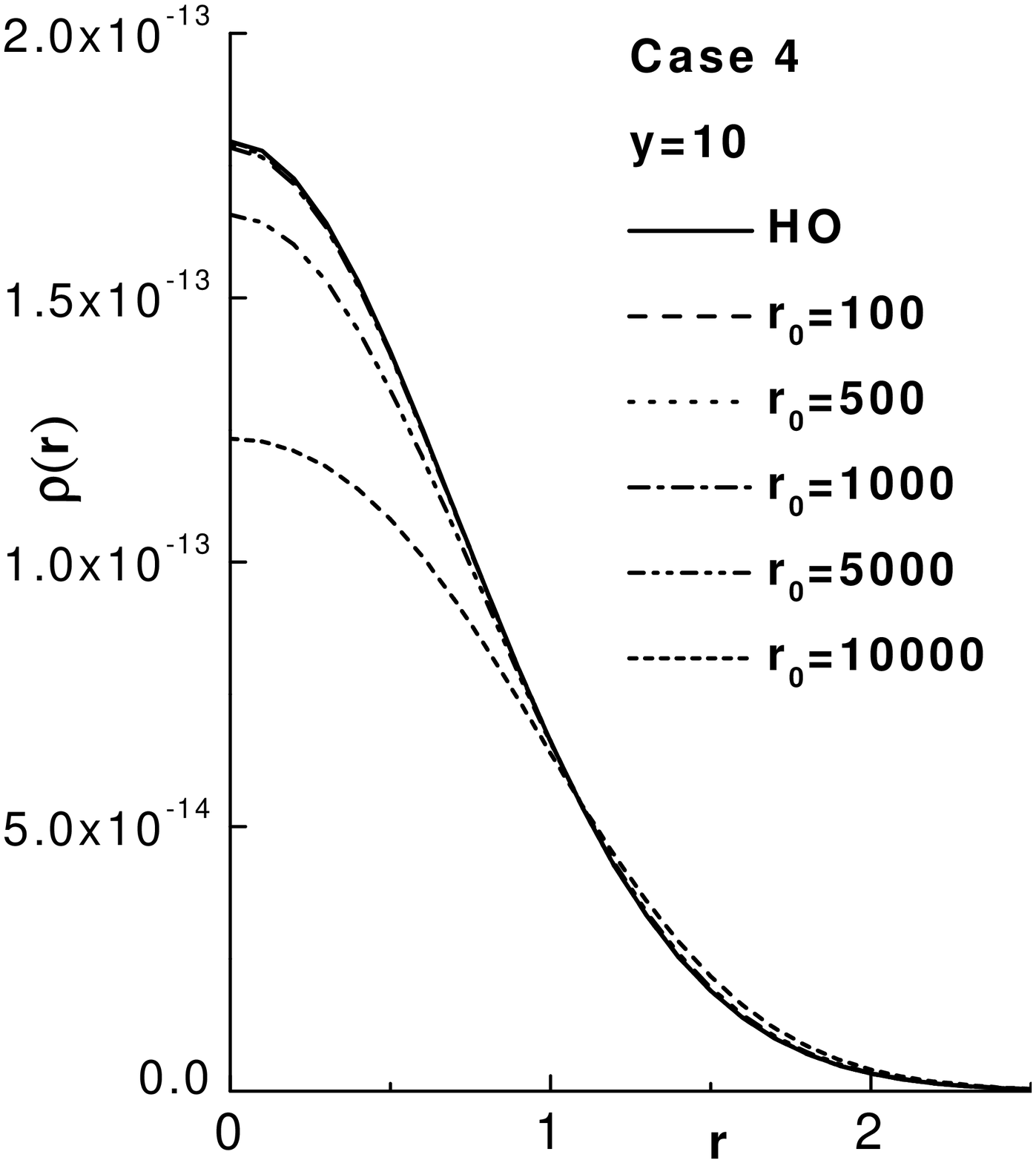,width=5.cm} }&
{\epsfig{figure=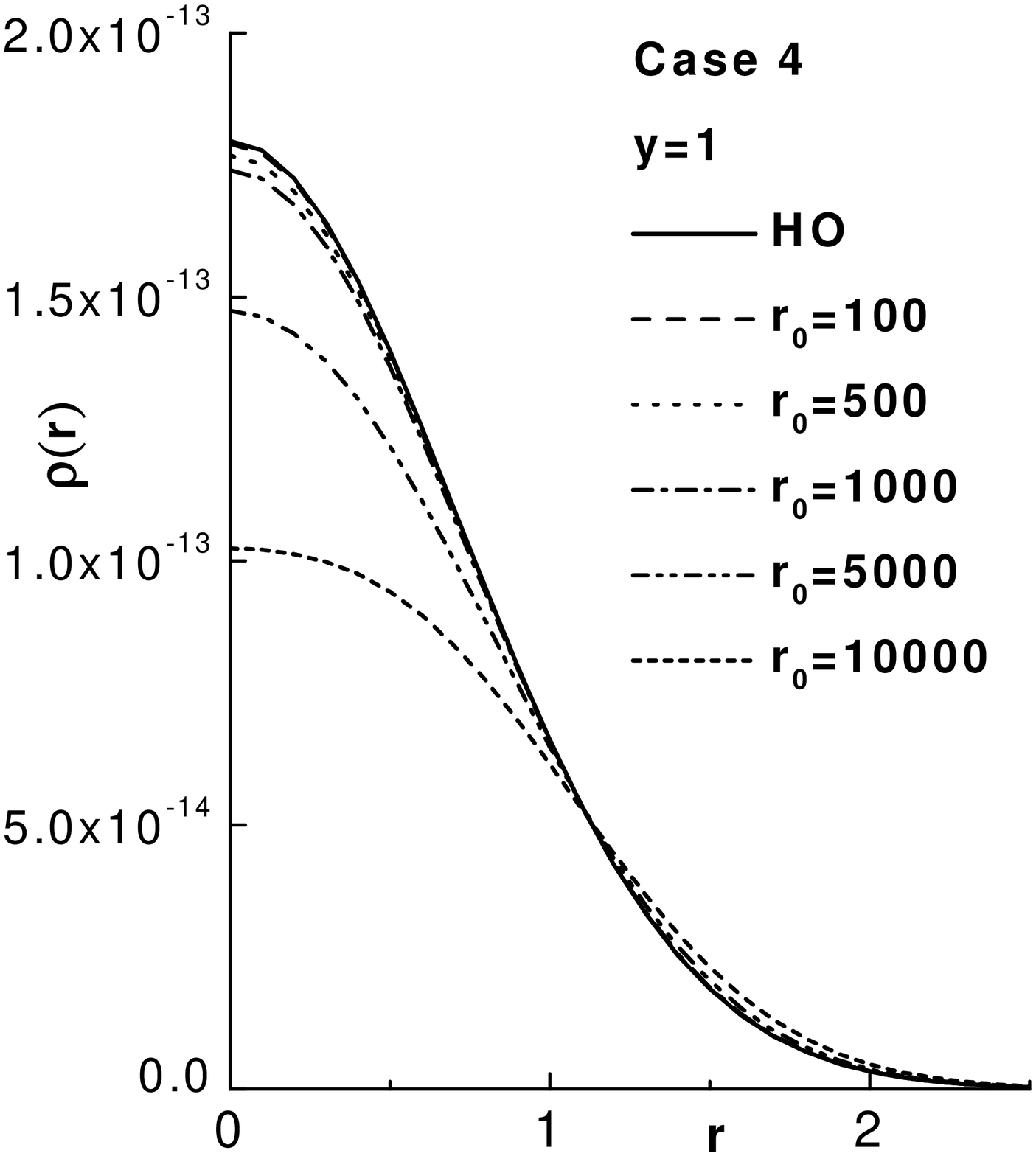,width=5.cm} }&
{\epsfig{figure=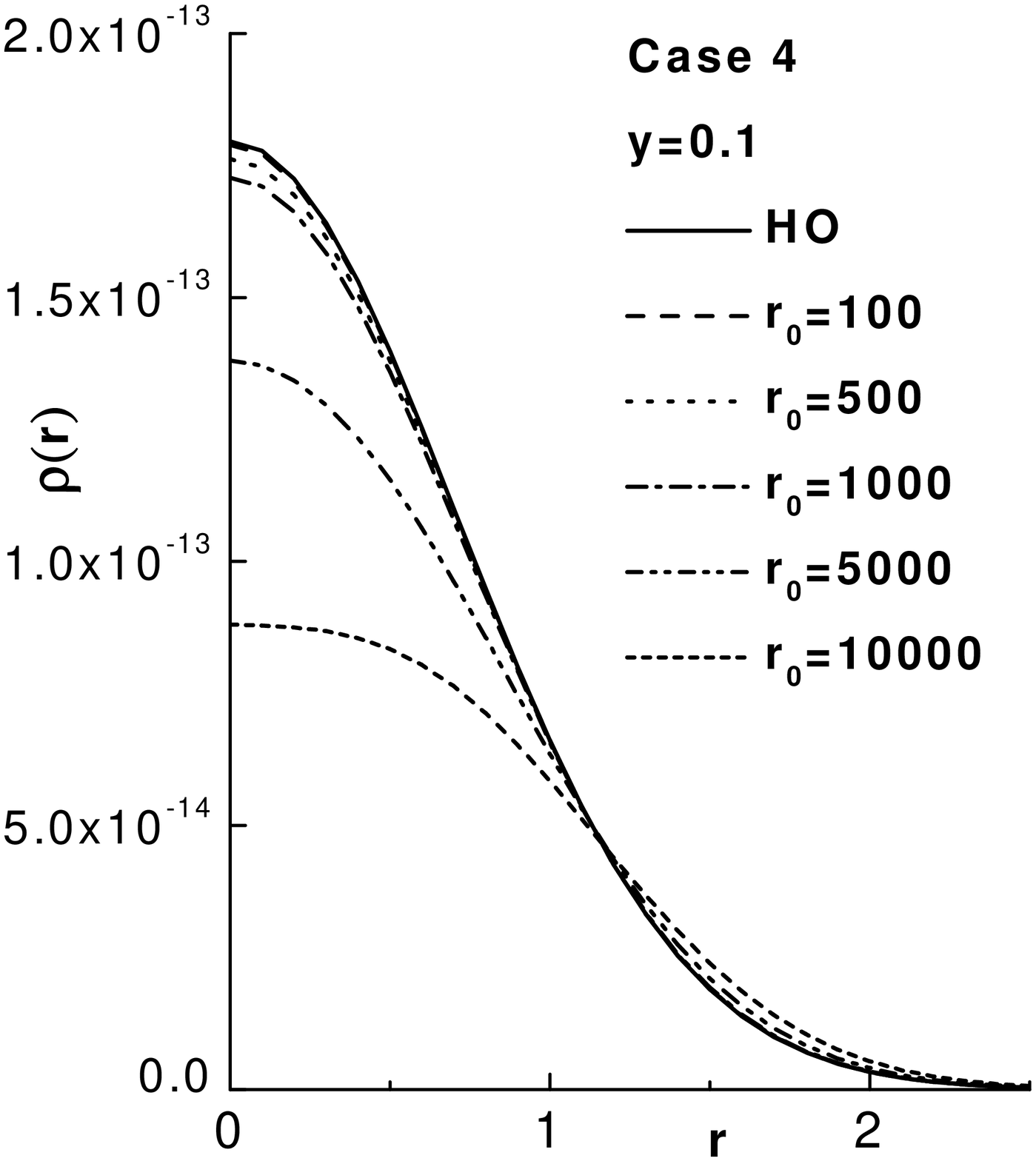,width=5.cm}}
\end{tabular}
\end{center}
\caption{The DD in case 4 versus $r$ for various values of the parameters
$y$ and $r_0$. The units and the normalization are as in Fig. 1.}
\end{figure}


\begin{figure}
\begin{center}
\begin{tabular}{lr}
{\epsfig{figure=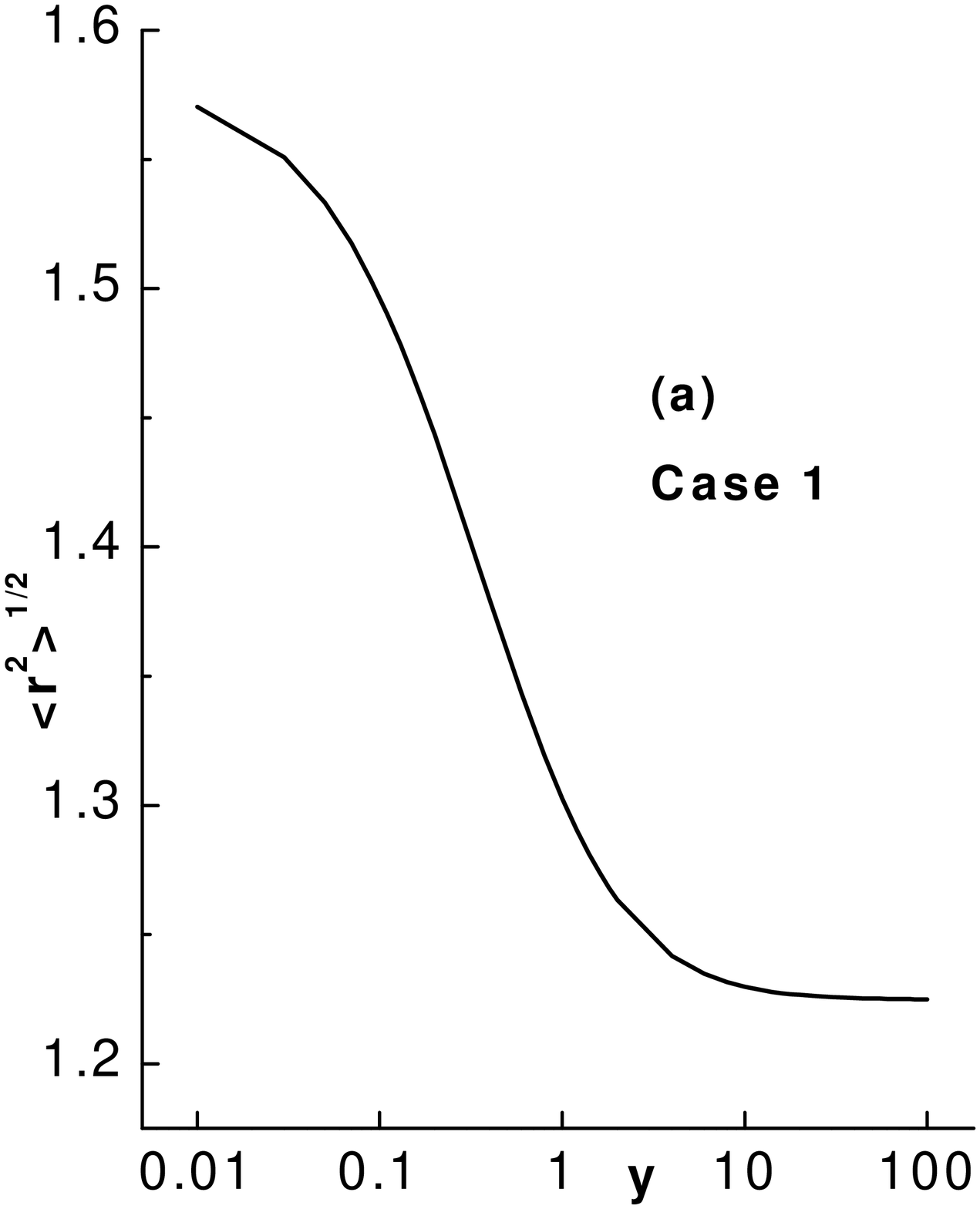,width=5cm} }&
{\epsfig{figure=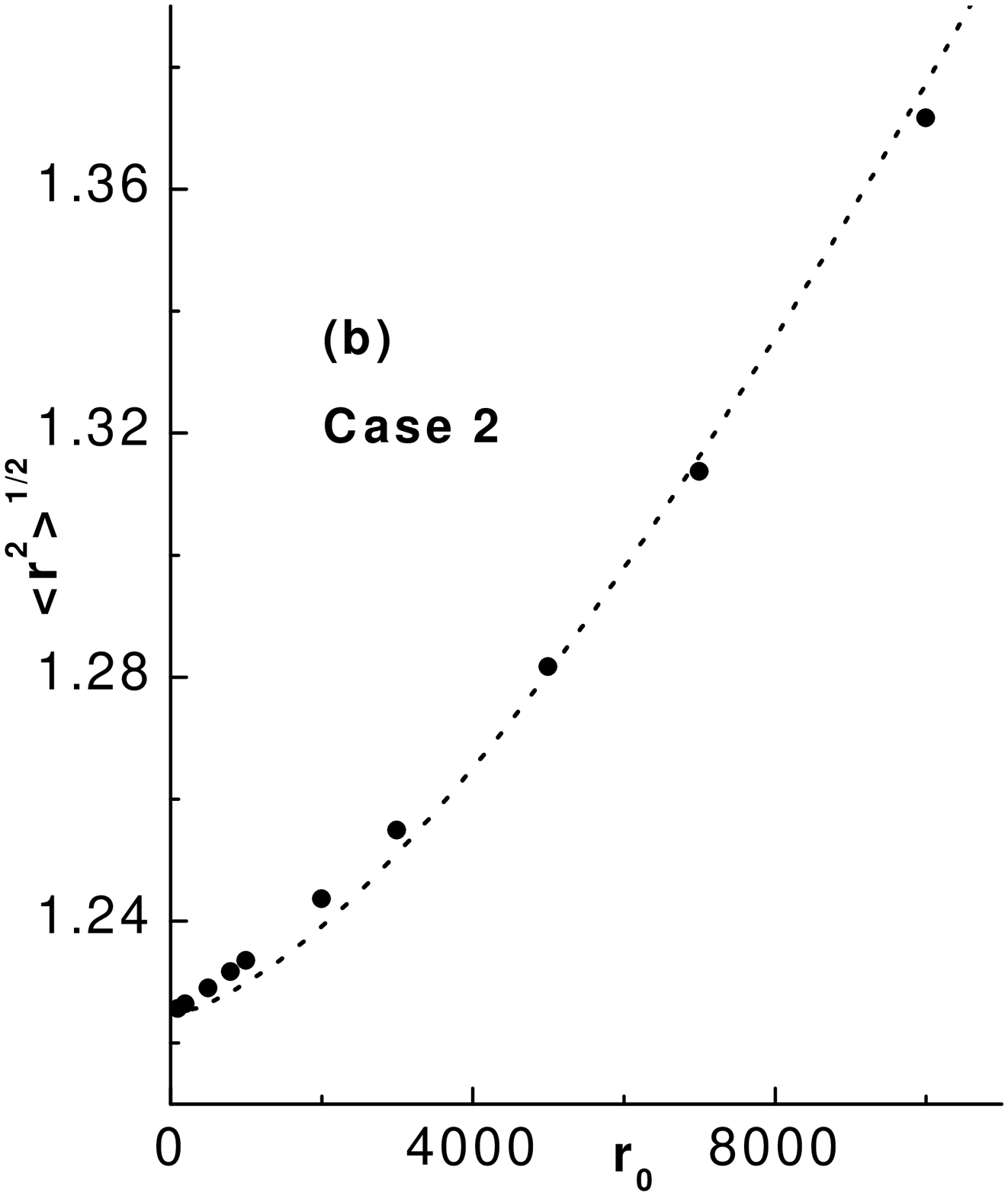,width=5cm} }\\
\end{tabular}
\end{center}
\caption{The rms radius, $\langle r ^{2} \rangle^{1/2}$, in cases 1
and 2 versus the parameters $y$ and $r_0$, respectively.
Case 1 corresponds to the analytical calculations of
$\langle r ^{2} \rangle^{1/2}$, while the points in case 2 correspond to
the numerical calculations of it. The dashed line in case 2 corresponds
to the fitting expression (50).
$\langle r ^{2} \rangle^{1/2}$ is in units of the trap lenght $b$. }
\vspace*{1cm}
\end{figure}

\begin{figure}
\begin{center}
\begin{tabular}{lr}
{\epsfig{figure=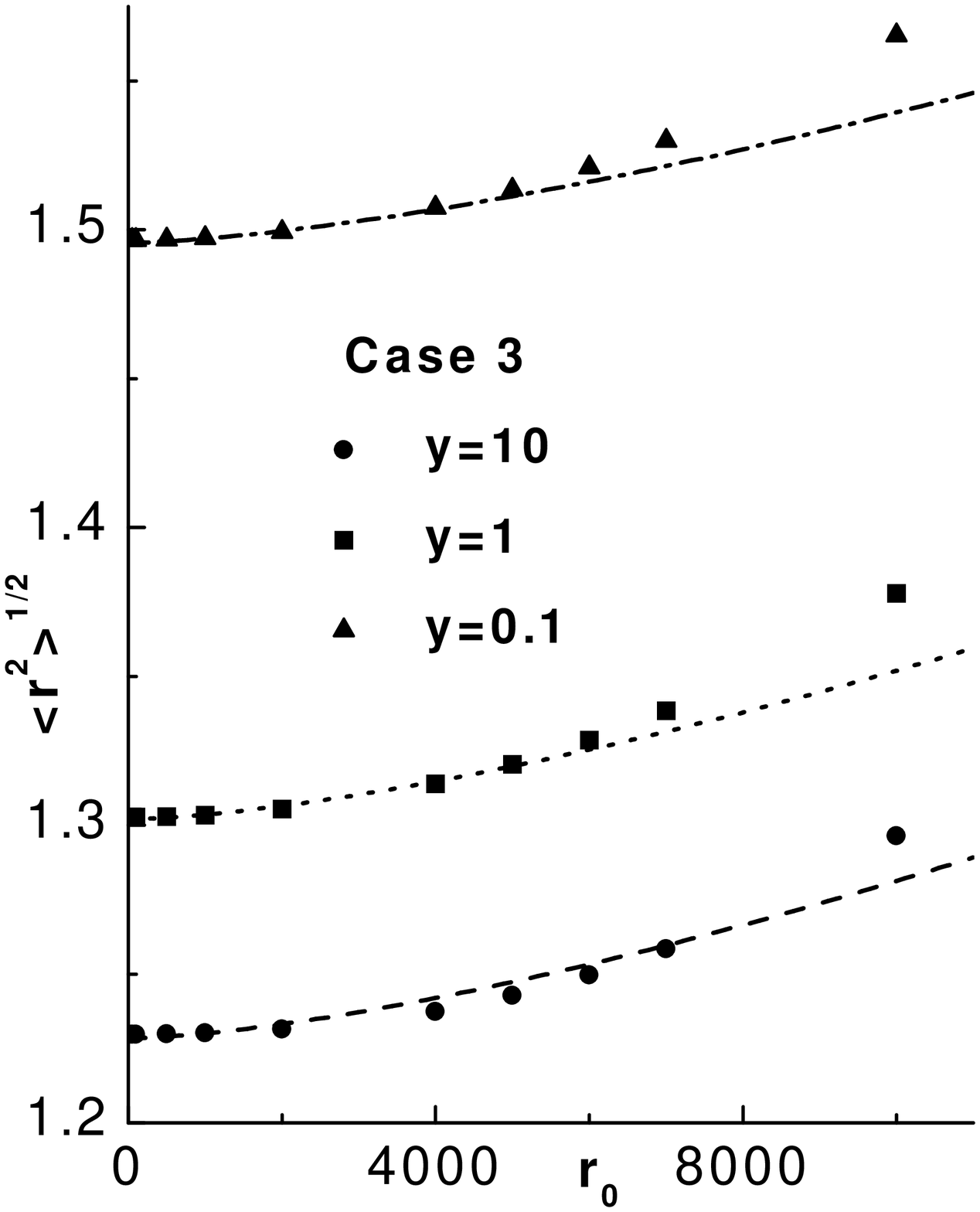,width=5cm} }&
{\epsfig{figure=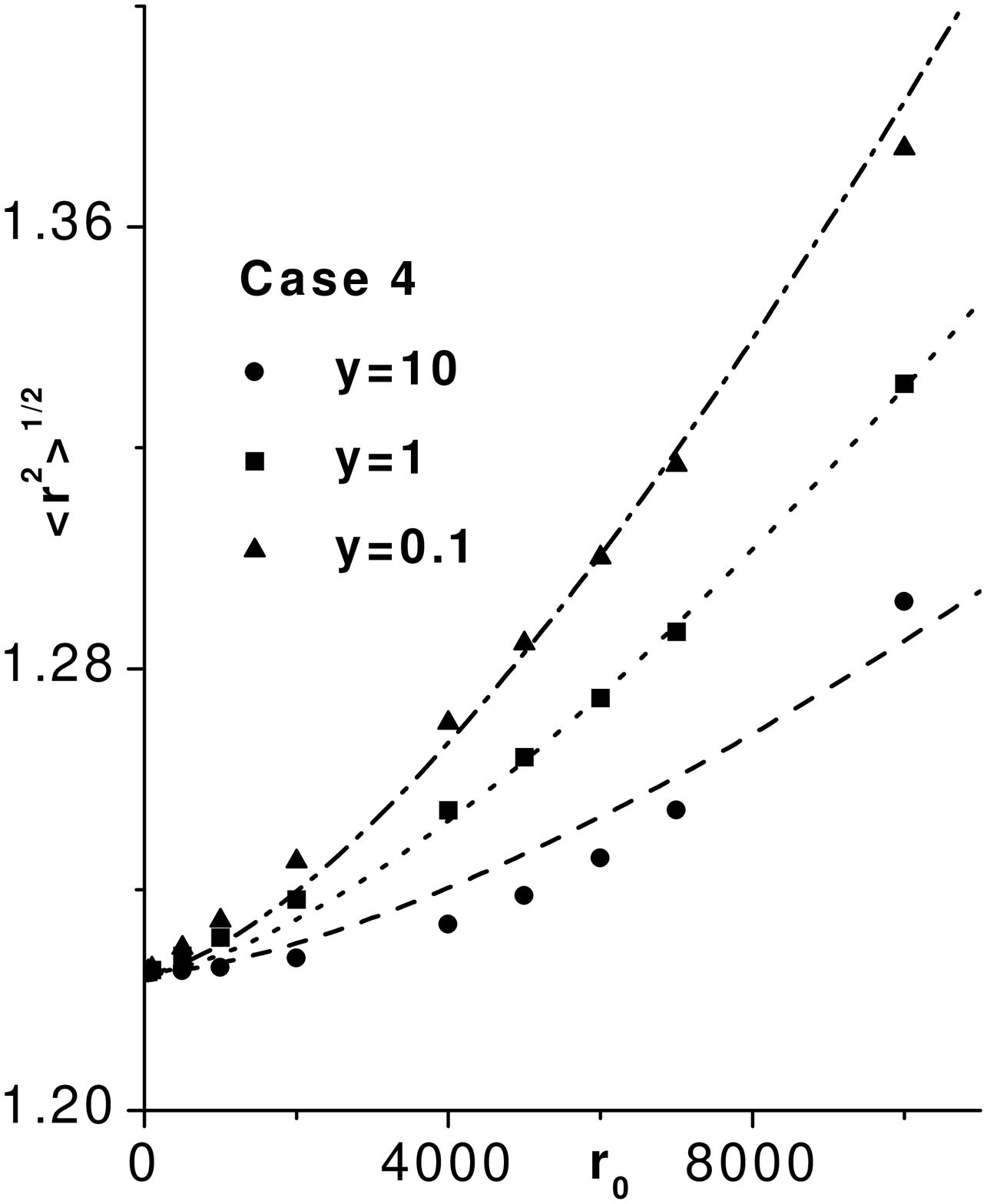,width=5cm} }\\
\end{tabular}
\end{center}
\caption{The rms radii in cases 3 and 4 versus the parameter $r_0$.
The circles, squares and triangles correspond to the approximate calculations
of $\langle r ^{2} \rangle^{1/2}$ for three different values of the
parameter $y$, while the dashed, dot and dashed-dot curves correspond to
the fitting expressions (51) and (52) for the same values of $y$.
$\langle r ^{2} \rangle^{1/2}$ is in units of the trap lenght $b$.}
\end{figure}

\newpage
\begin{figure}
\begin{center}
\begin{tabular}{ccc}
{\epsfig{figure=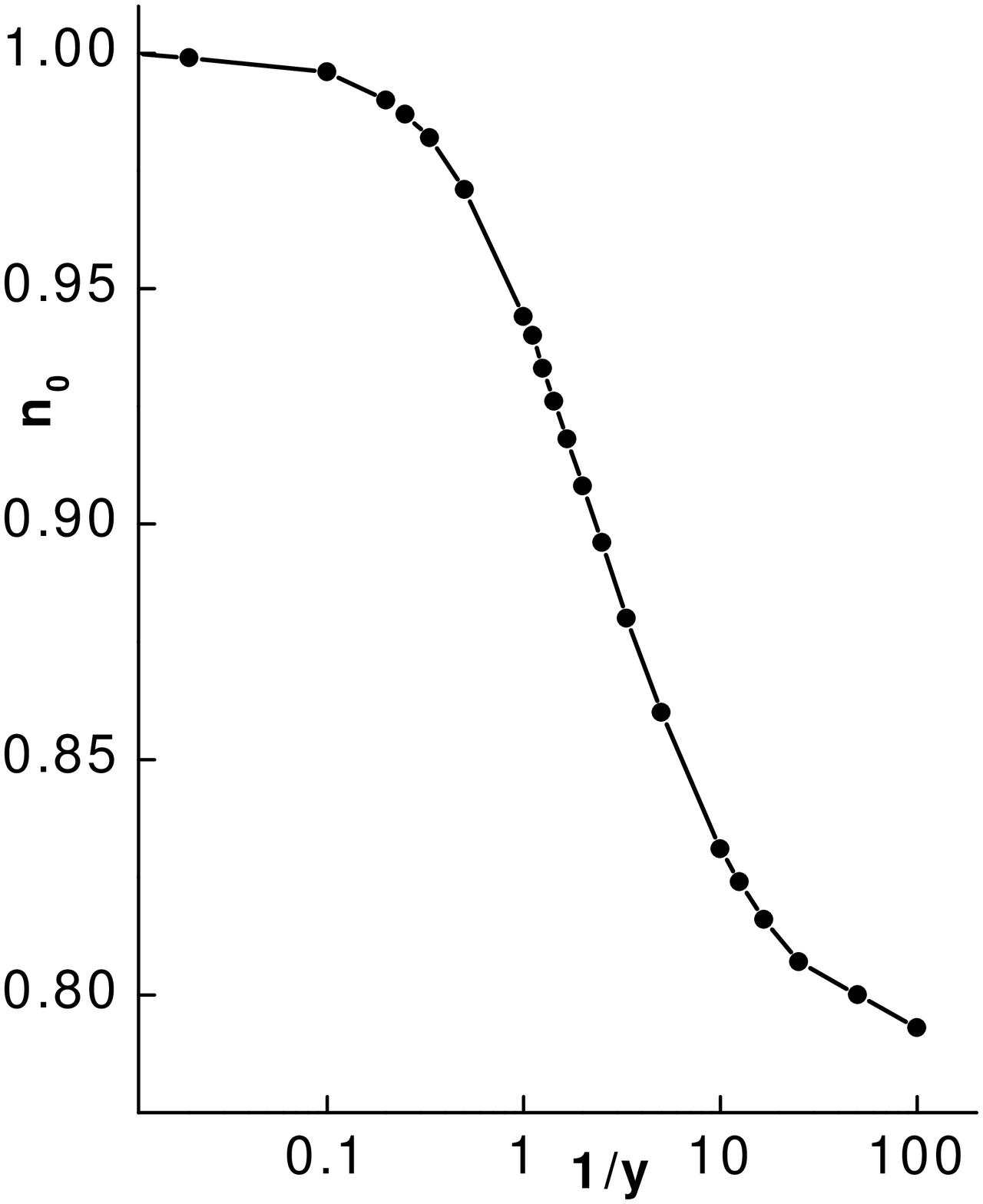,width=5.cm} }
\end{tabular}
\end{center}
\caption{ The condensate fraction $n_0$, at zero temperature, for
interacting atoms in case 1
versus the correlation parameter $1/y$.}
\end{figure}

\vspace*{1cm}
\begin{figure}
\begin{center}
\begin{tabular}{ccc}
{\epsfig{figure=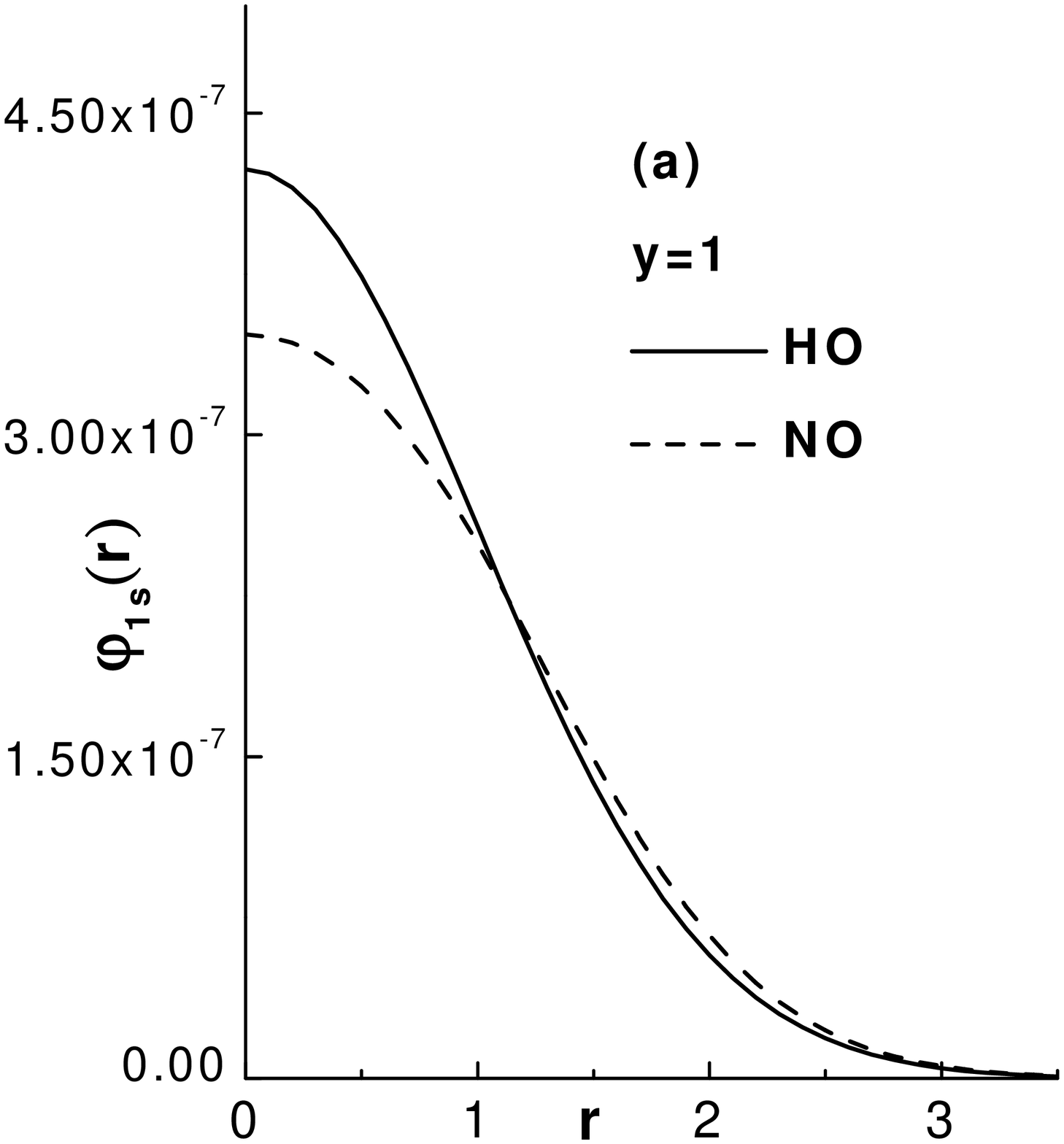,width=5.cm} }&
{\epsfig{figure=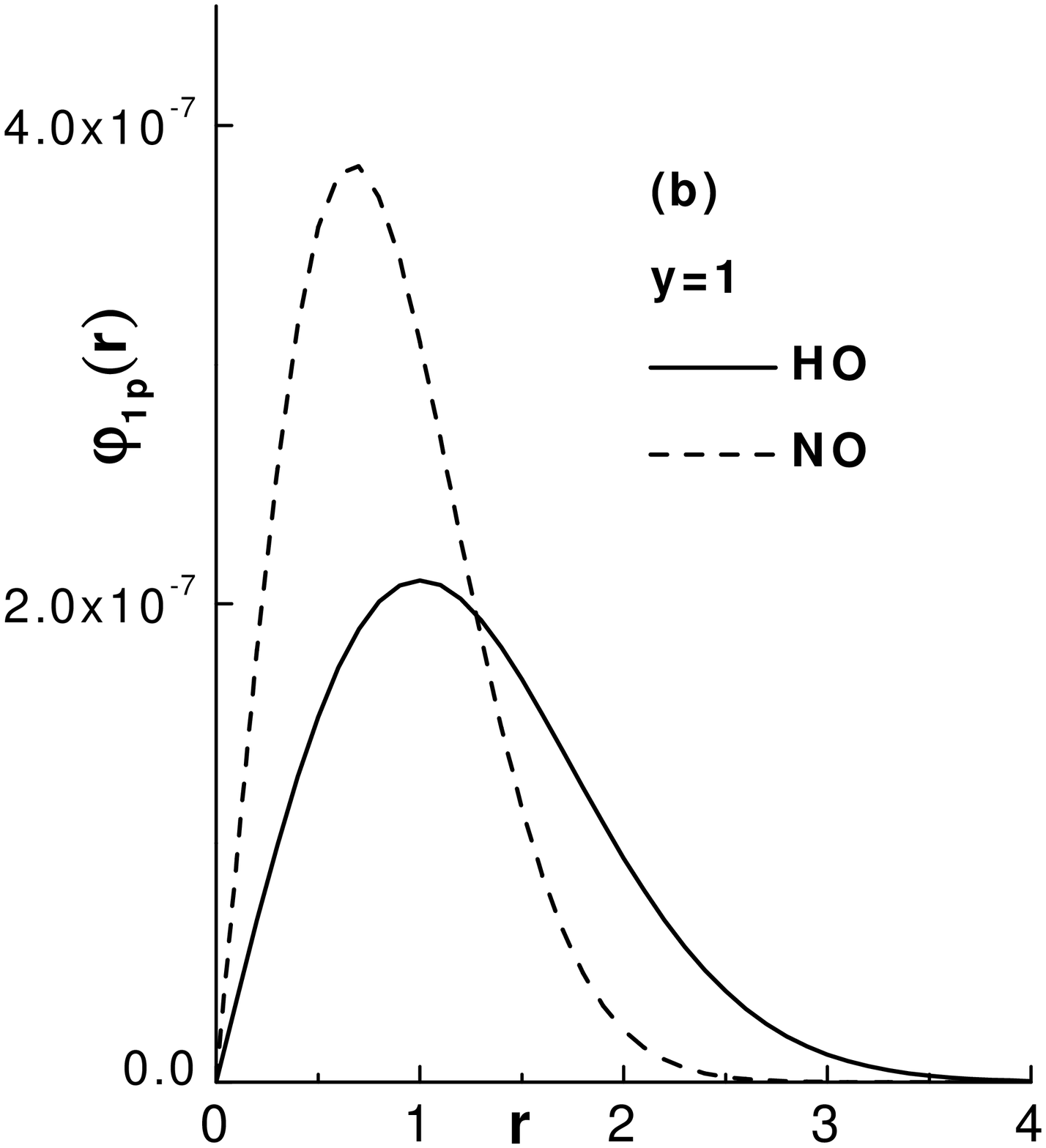,width=5.cm} }&
{\epsfig{figure=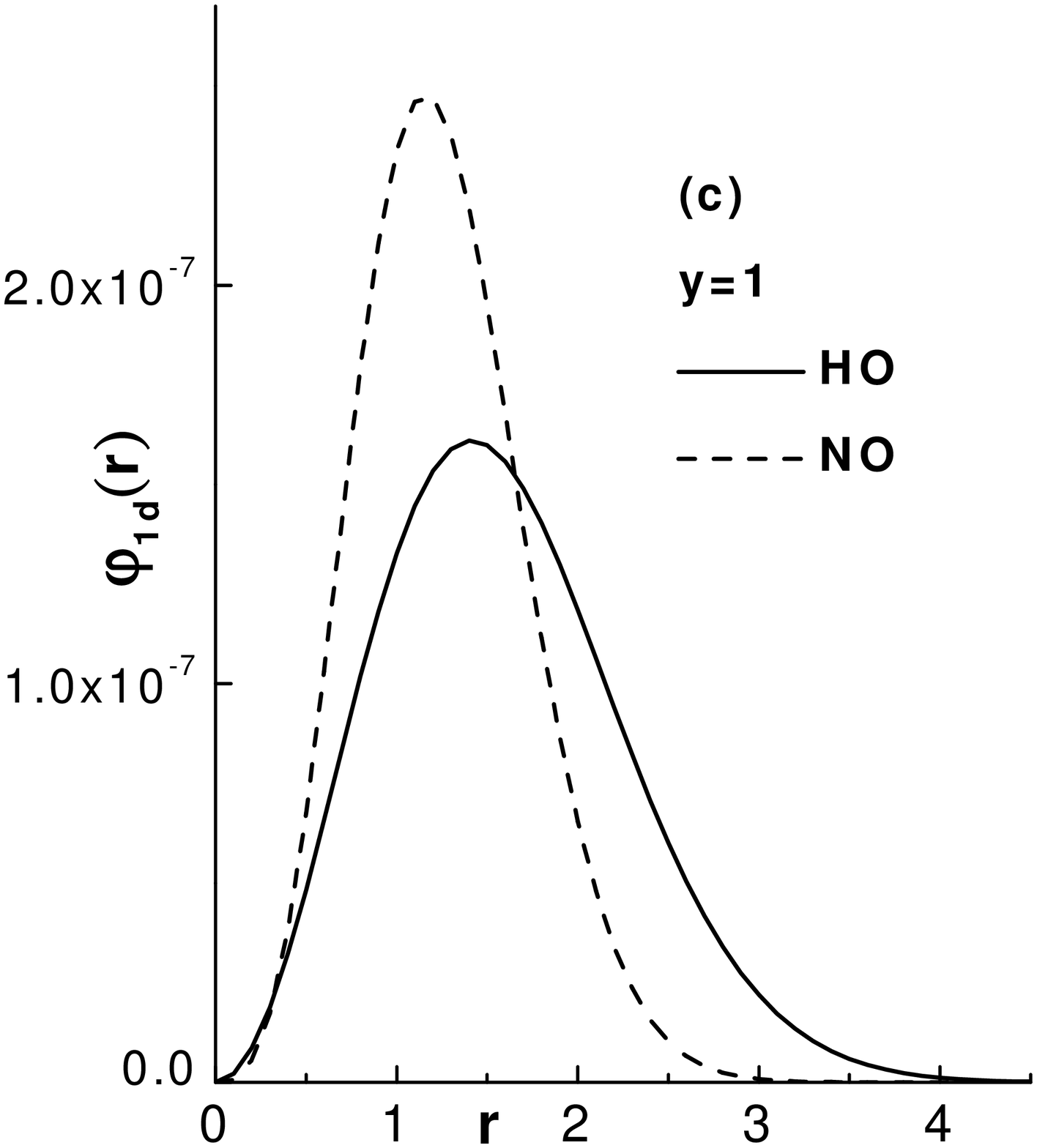,width=5.cm}}
\end{tabular}
\end{center}
\caption{The NO's (dashed line) of the states $1s$, $1p$ and $1d$
obtained by diagonalization of the OBDM
in case 1. The solid lines  correspond to the HO wave-function
with the trap length b.}
\end{figure}

\vspace*{1cm}

\newpage
\begin{figure}
\begin{center}
\begin{tabular}{lr}
{\epsfig{figure=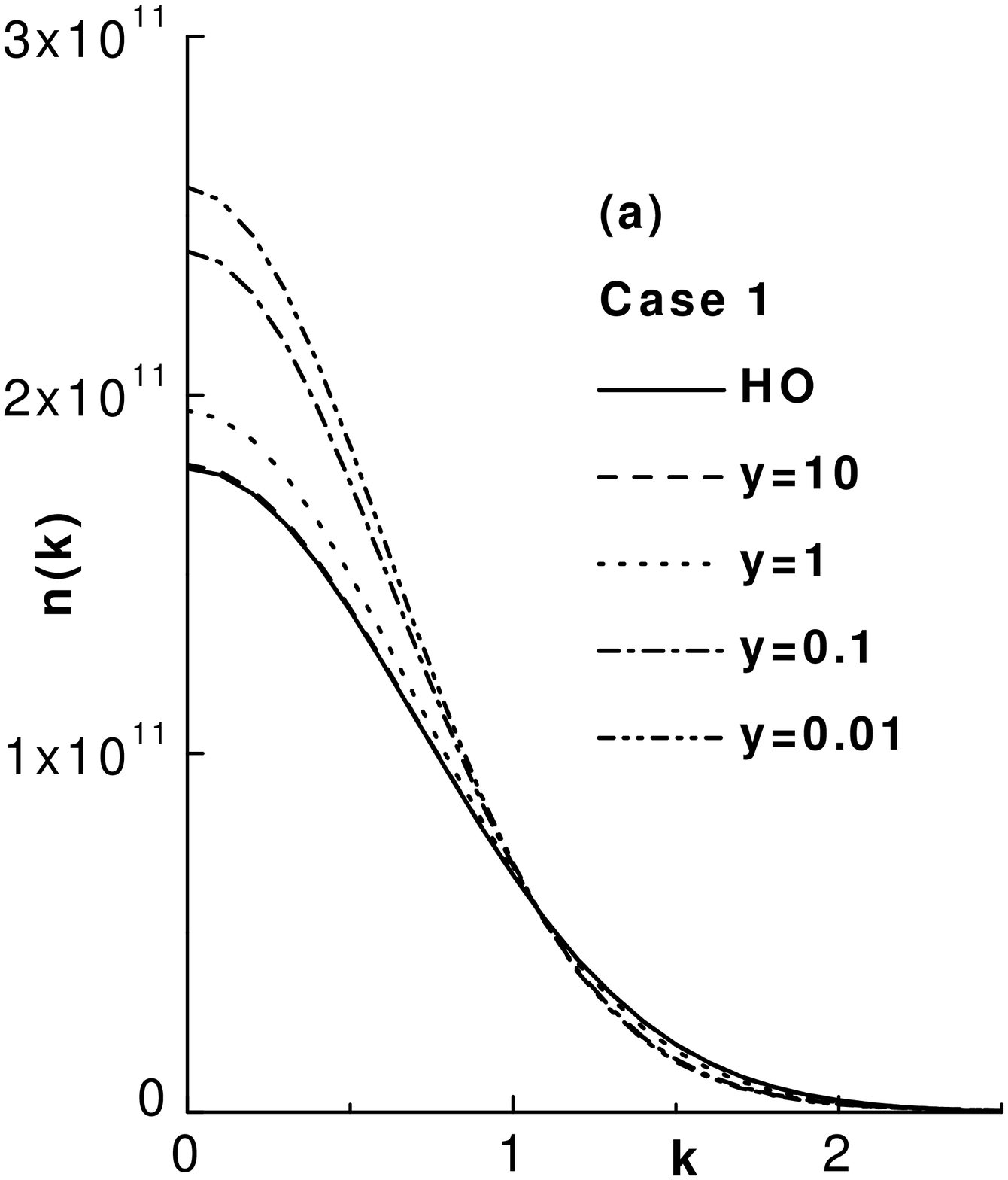,width=5cm} }&
{\epsfig{figure=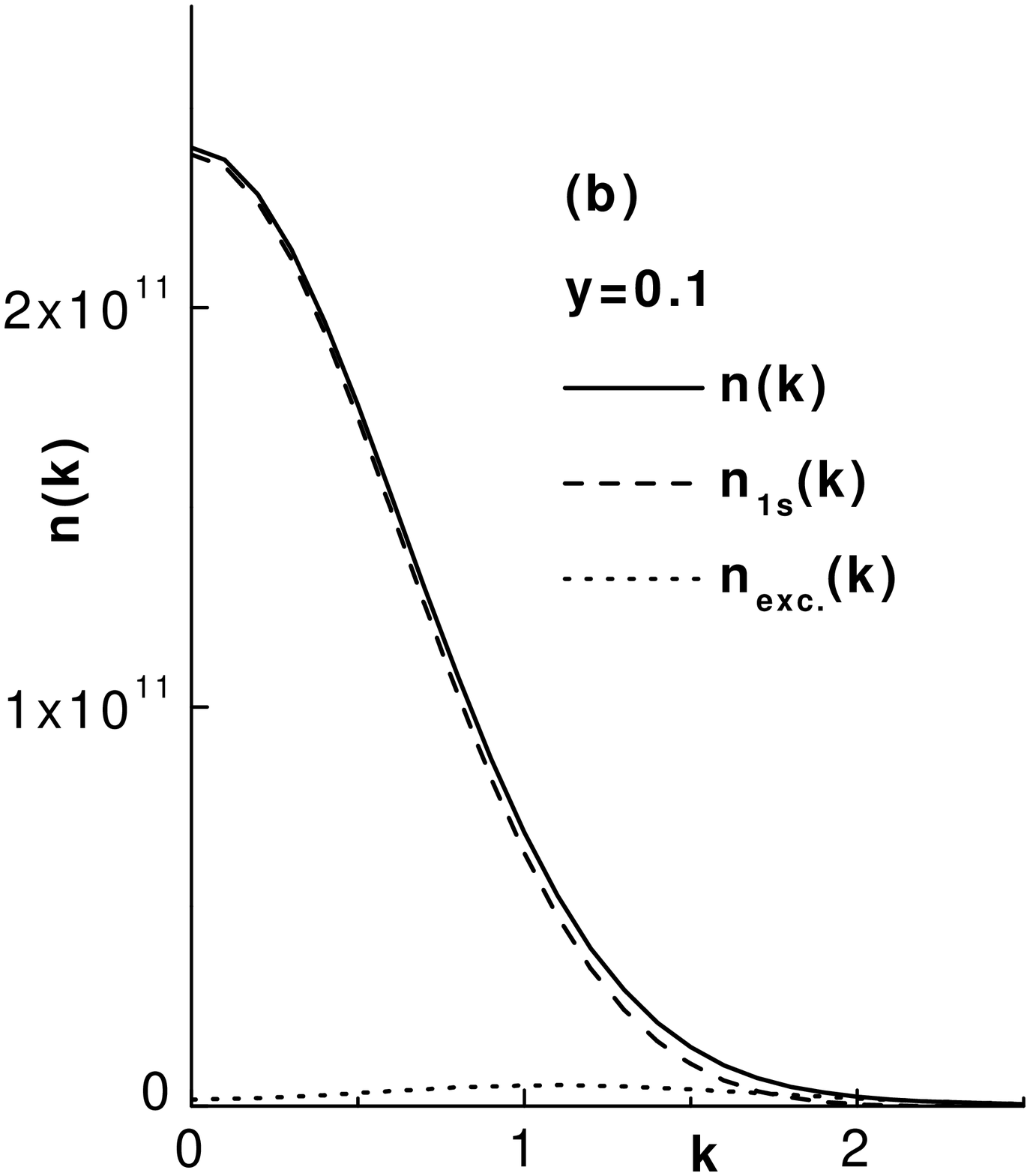,width=5cm} }\\
\end{tabular}
\end{center}
\caption{(a) The MD (normalized to $1$) in case 1 versus $k$
for various values of the parameters $y$. 
(b) The MD and the contribution to it from the NO's of the $1s$-state
and of all the excited states for $y=0.1$.
$k$ is in units of
the inverse of the trap length ($b^{-1}$) and  the MD's  are in $\AA^{3}$.}
\end{figure}

\vspace*{1cm}
\begin{figure}
\begin{center}
\begin{tabular}{lr}
{\epsfig{figure=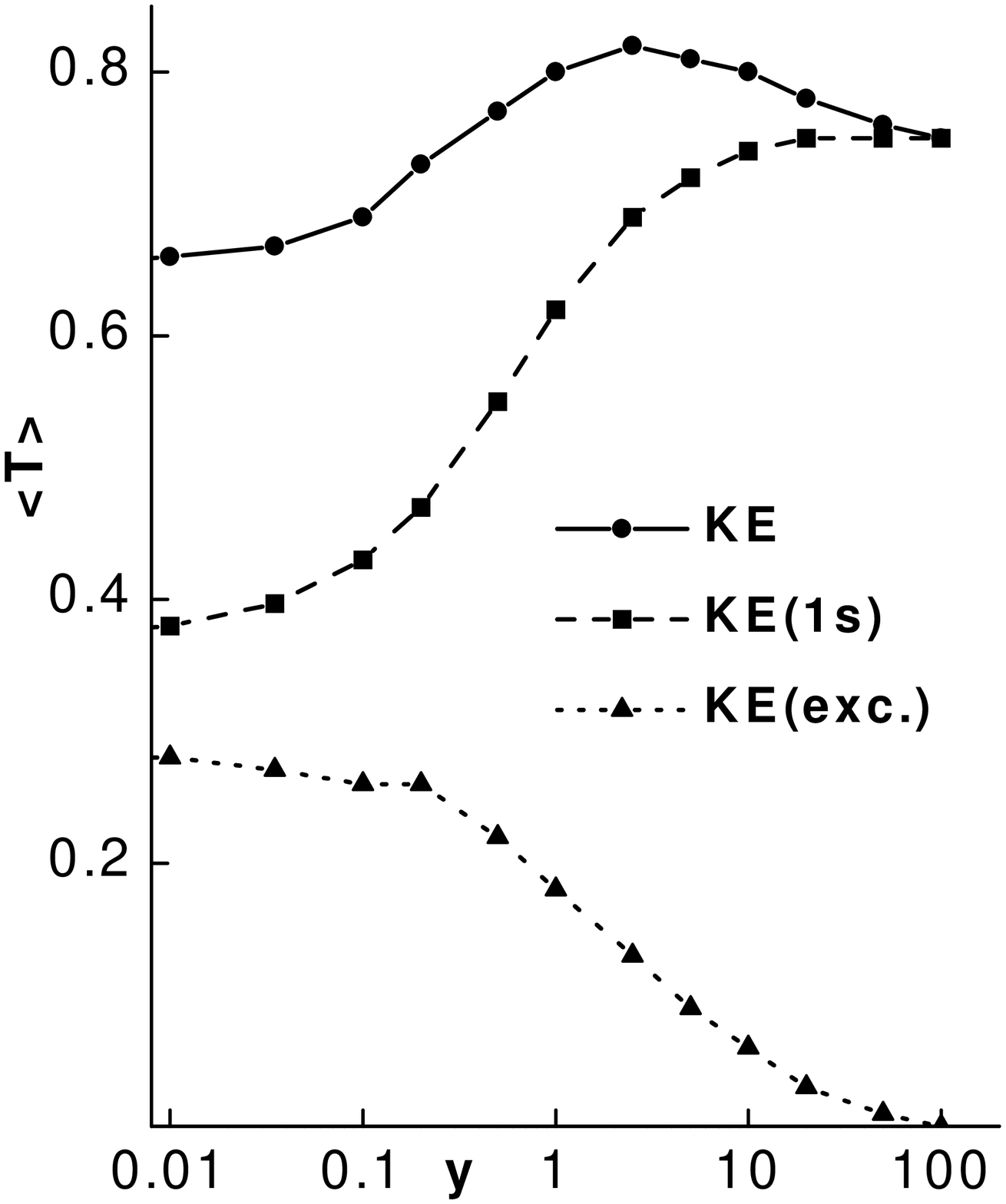,width=5cm} }
\end{tabular}
\end{center}
\caption{The mean kinetic energy per atom, $\langle T \rangle$, in case
1 versus the parameters $y$.
The solid curve corresponds to the total values
of $\langle T \rangle$, while the
dashed line and the dotted line
are the contributions to the
total $\langle T \rangle$ of the NO's of the $1s$-state
and of all the excited states,
respectively.}
\end{figure}

\end{document}